\documentclass[prb,twocolumn,showpacs,superscriptaddress,floatfix]{revtex4}
\usepackage{graphicx,amsfonts,amssymb,amsmath,hyperref}

\newif\ifhyper
\hypertrue
\ifhyper
\hypersetup{
  citecolor = {green},
  colorlinks = {true}, 
  urlcolor = {blue} 
}
\fi

\newlength{\ldag}
\newcommand{\rmi}{\mathrm{i}}

\newcommand{\bn}{{\boldsymbol{n}}}

\begin{document}

\title{Bound states in two-dimensional spin systems near the Ising
limit~: 
a quantum finite-lattice study}

\author{S\'ebastien Dusuel}
\email{sdusuel@gmail.com}
\affiliation{Lyc\'ee Saint-Louis, 44 Boulevard Saint-Michel, 75006 Paris, France}

\author{Michael Kamfor}
\email{kamfor@fkt.physik.uni-dortmund.de}
\affiliation{Lehrstuhl f\"ur Theoretische Physik I, Otto-Hahn-Stra\ss e 4, TU Dortmund, 44221 Dortmund, Germany}

\author{Kai Phillip Schmidt}
\email{schmidt@fkt.physik.uni-dortmund.de}
\affiliation{Lehrstuhl f\"ur Theoretische Physik I, Otto-Hahn-Stra\ss e 4, TU Dortmund, 44221 Dortmund, Germany}

\author{Ronny Thomale}
\email{rthomale@princeton.edu}
\affiliation{Lehrstuhl f\"ur Theoretische Physik I, Otto-Hahn-Stra\ss e 4, TU Dortmund, 44221 Dortmund, Germany}
\affiliation{Institut f\"{u}r Theorie der Kondensierten Materie, Universit\"{a}t Karlsruhe, D-76128 Karlsruhe, Germany}
\affiliation{Department of Physics, Princeton University, Princeton, New Jersey 08544, USA}

\author{Julien Vidal}
\email{vidal@lptmc.jussieu.fr}
\affiliation{Laboratoire de Physique Th\'eorique de la Mati\`ere Condens\'ee,
CNRS UMR 7600, Universit\'e Pierre et Marie Curie, 4 Place Jussieu, 75252
Paris Cedex 05, France}


\begin{abstract}
  We analyze the properties of low-energy bound states in the transverse-field Ising
  model and in the XXZ model on the square lattice. To this end, we develop an optimized  implementation of perturbative continuous unitary transformations. The Ising model is
  studied in the small-field limit which is found to be a special case of the toric code
  model in a magnetic field. To analyze the XXZ model, we perform a perturbative expansion about the Ising limit 
  in order to discuss the fate of the elementary magnon excitations when approaching the Heisenberg point. 
\end{abstract}


\maketitle

\section{Introduction}

Properties of strongly correlated quantum matter are usually well described 
by the many-body ground state and by the first elementary excitation. Multiparticle 
excitations are often not important because they just constitute an incoherent background.
Thus, the study of quantum phase transitions mainly relies on low-energy spectrum analysis, 
namely energy levels of the ground state and the first excited state \cite{Sachdev99}. 
However, in various systems, bound states may arise and play a major role. One of the most famous example are Cooper pairs which lead to superconductivity but other electron systems also display such mechanism  (see, for instance, Refs.~\onlinecite{Doucot02,Vidal00_1,Capponi07}).
In one-dimensional magnetic systems, well-known examples are the dimerized 
and frustrated spin chain as well as the two-leg spin ladder \cite{Uhrig96,Trebst00,Zheng01,Knetter01_2,Zheng03,Schmidt04} which contains bound states made up of triplons. Interestingly, such modes have been experimentally
observed in cuprate ladder materials \cite{Windt01}. 
In two dimensions, the frustrated Shastry-Sutherland model and its experimental realization SrCu$_2$(BO$_3$)$_2$ are also known to possess two-triplon bound states \cite{Knetter00_2,Lemmens00}.
Note that more complicated bound states may arise in topologically ordered systems where anyons (semions) can bind to form bosons or fermions as discussed in Ref.~\onlinecite{Vidal09_2}.

Recently, such binding effects have been studied in the two-dimensional XXZ model \cite{Hamer09} where elementary excitations are dressed magnons. 
 The aim of the present paper is to analyze the spectrum of such magnon bound states in two different spin systems.  
The first one is the ferromagnetic transverse-field Ising model (TFIM) on the square lattice for which we derive the two-magnon spectrum perturbatively in the small-field limit. The high-order series expansion (order 12) of the corresponding gap allows us to extrapolate its behavior near the critical point where it is found to vanish.  We also compute  the ratio between the one-magnon gap and this two-magnon gap at the critical point which is approximately $1.8$, in agreement with field-theoretical predictions \cite{Caselle99,Caselle02}.
The second system considered in this study is the XXZ model on the square lattice which, in the isotropic limit (XXX), is the celebrated antiferromagnetic Heisenberg model. As for the TFIM, we focus on the two-magnon bound-states spectrum which is derived up to order 8 near the Ising limit.
However, as we shall see, results obtained here differ from those obtained recently by Hamer \cite{Hamer09} since, contrary to his claim, we show using very simple arguments,  that there are two distinct branches of two-magnon bound states at low energy. Also, let us underline that the fate of the lowest-energy bound state when approaching the Heisenberg limit cannot be conclusively determined  at this order. 

From a methodological point of view, several methods to compute
high-order series expansion in quantum many-body systems are available
(see Refs.~\onlinecite{Gelfand00,Oitmaa06} for a review). Here, we use the
perturbative continuous unitary transformations (PCUTs) method 
\cite{Wegner94,Stein97,Knetter00_1,Knetter03_1} which is especially well
suited to investigate the many-particle spectrum and provides a
natural quasiparticle (QP) \mbox{description}. 

The structure of this paper is the following.
In Sec.~\ref{sec:models_methods}, we introduce the two models (TFIM in Sec.~\ref{sec:sub:TFIM} and XXZ  model in Sec.~\ref{sec:sub:Heisenberg_XXZ}) under consideration. We show that the Ising limit is a good starting point for a perturbation theory in the PCUTs framework and we also give a very simple picture to understand the occurrence of bound states in this limiting case. In Sec.~\ref{sec:sub:bond_description}, we introduce another description of the spin model in terms of bond degrees of freedom which will be useful to set up the present perturbation theory framework. The end of this Sec.~\ref{sec:sub:symm_count} is dedicated to symmetry considerations. 
In Sec.~\ref{sec:PCUT}, we recall several important aspects of PCUTs which are essential to understand the next section. In Sec.~\ref{sec:LCE_QFL},  we adapt the finite-lattice method (commonly used in statistical mechanics 
\cite{Enting96,Oitmaa06}) to quantum problems, allowing one to significantly increase the maximum order of the series expansions. Let us stress that readers interested only in results can skip these two sections and switch directly to Secs.~\ref{sec:results_TFIM} and \ref{sec:results_XXZ} where we discuss the low-energy spectra of the TFIM and XXZ model, respectively. Finally, in Sec.~\ref{sec:TC},  we discuss the spectrum of a new model which naturally emerges when describing the TFIM in terms of bonds. This model may be seen as a special case of the toric code model \cite{Kitaev03} in a magnetic field \cite{Vidal09_1,Vidal09_2} in which flux creation energy cost vanishes. 
All coefficients used to compute gaps in both models are gathered in Appendices A and B.

\section{Models and mappings}
\label{sec:models_methods}
 
\subsection{Transverse-field Ising model}
\label{sec:sub:TFIM}

%
\begin{figure}[t]
  \includegraphics[width=0.6\columnwidth ]{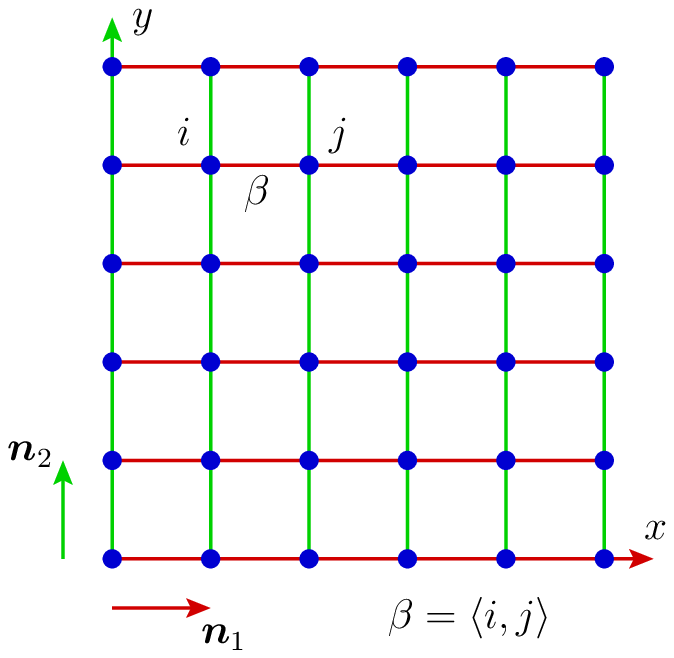}
  \caption{ (Color online)  Square lattice on which the models studied
    in this paper are defined. Translation vectors in the $x$ and $y$
    directions are denoted $\bn_1$ and $\bn_2$. The bond between sites
    $i$ and $j$ is denoted $\langle i, j\rangle$ or by a Greek letter
    such as $\beta$ to shorten the notation.  }
  \label{fig:lattice}
\end{figure}
%

Let us first introduce the TFIM on the
square lattice whose Hamiltonian reads
%
\begin{equation}
  \label{eq:hamTFIM}
  H_\mathrm{TFIM} = -J \sum_{\langle i,j \rangle} \sigma_i^z \sigma_j^z
  - h \sum_{i} \sigma_i^x,
\end{equation} 
%
where the second sum runs over all  $i$ of the square lattice and the
first sum runs over all bonds $\langle i,j \rangle$ between nearest-neighbor sites (see Fig.~\ref{fig:lattice}). The Pauli matrices are
denoted by $\sigma^x$, $\sigma^y$, and $\sigma^z$.  Here, we focus on
a ferromagnetic Ising coupling $J>0$ and, without loss of generality,
we consider $h \geq 0$.

This model whose classical counterpart is the three-dimensional Ising
model is known to display a second-order phase transition separating a
symmetric (disordered) phase at large field $h$ from a broken
(ordered) phase at large coupling $J$. For an infinite field, all
spins point in the $+x$ direction, while for an infinite exchange
coupling, all spins point either in the $+z$ direction or in the $-z$
direction.

The ground-state energy as well as the single-excitation spectrum have
been computed in both phases using series expansion
\cite{He90,Oitmaa91,Oitmaa06} allowing for a precise determination of
the critical point at $J/h|_c=0.3285(1)$. In the broken phase, the
perturbative expansion is done around the Ising limit ($h=0$). Thus, it
is convenient to introduce the following Hamiltonian that will also be
the starting point of our study of the Heisenberg model (see Sec.~\ref{sec:sub:Heisenberg_XXZ}) 
%
\begin{equation}
  \label{eq:hamI}
  H_\mathrm{I} = -\frac{1}{2} \sum_{\langle i,j \rangle} \sigma_i^z \sigma_j^z.
\end{equation} 
%

\begin{figure}[t]
  \includegraphics[width=0.6\columnwidth  ]{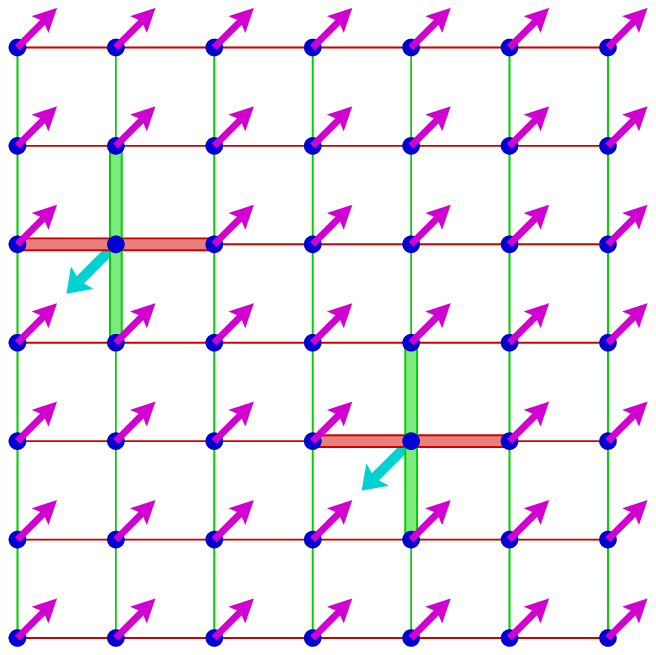}
  \caption{
  (Color online)  Two magnons (spins pointing left) on top of a ferromagnetic
  background. Such a state involves eight antiferromagnetic (bold) bonds
  (four for each magnon), and costs an energy $8$, with respect to Hamiltonian
  $H_\mathrm{I}$.
}
  \label{fig:ferro_2flip_far}
\end{figure}

Let us consider its ferromagnetic ground state where all spins point
in the $+z$ direction. Lowest-energy excitations then consist in
static magnons (spin flips) whose energy cost is $4$. The prefactor
$1/2$ in Eq.~(\ref{eq:hamI}) defining $H_\mathrm{I}$ indeed ensures
that any state will have an energy equal to that of the ground state
plus the number of antiferromagnetic bonds. Two such ``isolated" magnons are shown in
Fig.~\ref{fig:ferro_2flip_far}.

%
%

\begin{figure}[t]
  \includegraphics[width=0.6\columnwidth  ]{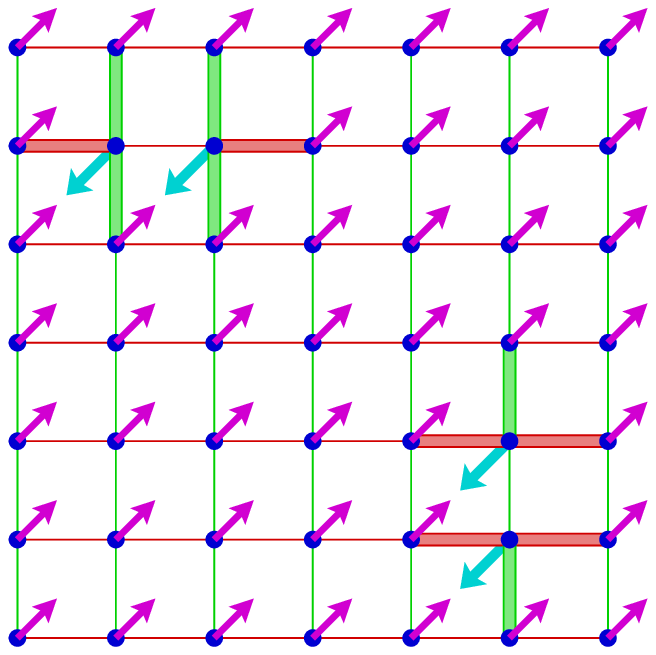}
  \caption{ (Color online)  Two examples of two-magnon bound states. Each of them
    has an energy cost of $6$, with respect to Hamiltonian
    $H_\mathrm{I}$. As in Fig.~\ref{fig:ferro_2flip_far},
    antiferromagnetic bonds are represented as bold segments.  }
  \label{fig:ferro_2flip_close}
\end{figure}

A two-magnon state will have an energy cost of $8$, except if the two
magnons are nearest neighbors. Indeed, in such a case, only six bonds
have an antiferromagnetic configuration, which results in an energy
reduction of $8-6=2$ compared to the situation where magnons are far
apart. This proves the existence of bound states in the spectrum, two of
which are shown in Fig.~\ref{fig:ferro_2flip_close}.

%
%

Of course, following the same line of reasoning, one can show that
$n$-magnon bound states exist for any $n\geqslant 2$, but we shall
restrict our study to the case $n=2$. Let us also notice that in the
ordered phase, the perturbation is performed around the field term of
Eq.~(\ref{eq:hamTFIM}), which basically counts spin flips, so that no
binding effect is present in this limit.

\subsection{Heisenberg and XXZ models}
\label{sec:sub:Heisenberg_XXZ}

Let us now discuss, in the same spirit, the antiferromagnetic
Heisenberg model on the square lattice. Its Hamiltonian reads
%
\begin{equation}
  \label{eq:hamH}
  H_\mathrm{H} = J \sum_{\langle i,j \rangle}
  \boldsymbol{S}_i\cdot\boldsymbol{S}_j,
\end{equation} 
%
where $\boldsymbol{S}_i=\boldsymbol{\sigma}_i/2$ is the spin operator
at site $i$. In the following, we set $J=2$  and we introduce an anisotropy parameter $\lambda$ with the
aim of performing series expansion in this parameter, as was done in
Refs.~\onlinecite{Singh89} and \onlinecite{Zheng91,Oitmaa94,Oitmaa06}.  The Heisenberg
Hamiltonian is then the $\lambda=1$ limit of the following XXZ
Hamiltonian
%
\begin{eqnarray}
  \label{eq:hamXXZ_1}
  H_\mathrm{XXZ} &=& \frac{1}{2} \sum_{\langle i,j \rangle}
  \sigma_i^z \sigma_j^z
  + \frac{\lambda}{2} \sum_{\langle i,j \rangle}
  \left( \sigma_i^x \sigma_j^x + \sigma_i^y \sigma_j^y \right),\\
  \label{eq:hamXXZ_2}
  &=& \frac{1}{2} \sum_{\langle i,j \rangle}
  \sigma_i^z \sigma_j^z
  + \lambda \sum_{\langle i,j \rangle}
  \left( \sigma_i^+ \sigma_j^- + \sigma_i^- \sigma_j^+ \right),
\end{eqnarray} 
%
where we have introduced the usual raising and lowering operators
$\sigma_i^\pm=\frac{1}{2}\left(\sigma_i^x\pm\rmi \, \sigma_i^y\right)$. The
square lattice being bipartite, it is possible to perform a
rotation of angle $\pi$ around the $x$ axis on one of the
sublattices, namely $\left(\sigma^x,\sigma^y,\sigma^z\right)\to
\left(\sigma^x,-\sigma^y,-\sigma^z\right)$. This transformation leads,
without changing notations for the Pauli operators and making use of
$H_\mathrm{I}$ defined previously, to
%
\begin{equation}
  \label{eq:hamXXZ_3}
  H_\mathrm{XXZ} = H_\mathrm{I} + \lambda \sum_{\langle i,j \rangle}
  \left( \sigma_i^+ \sigma_j^+ + \sigma_i^- \sigma_j^- \right).
\end{equation}
%

We thus see that in the limit of a vanishing anisotropy parameter $\lambda$,
$H_\mathrm{XXZ}$ reduces to $H_\mathrm{I}$ so that the binding effects
discussed above for $H_\mathrm{TFIM}$ are also at work here.  The difference
 between both models is the perturbation term added to
$H_\mathrm{I}$.

Let us finally mention that from the point of view of the XXZ model, a
second-order quantum phase transition occurs at the Heisenberg point
$\lambda=1$, separating a gapped phase for $0<\lambda<1$ with a
twofold-degenerate ground state with N\'eel order from a gapless
phase with O(2) symmetry for $\lambda>1$.

\subsection{Bond description}
\label{sec:sub:bond_description}

%
\begin{figure}[t]
  \includegraphics[width=0.6\columnwidth  ]{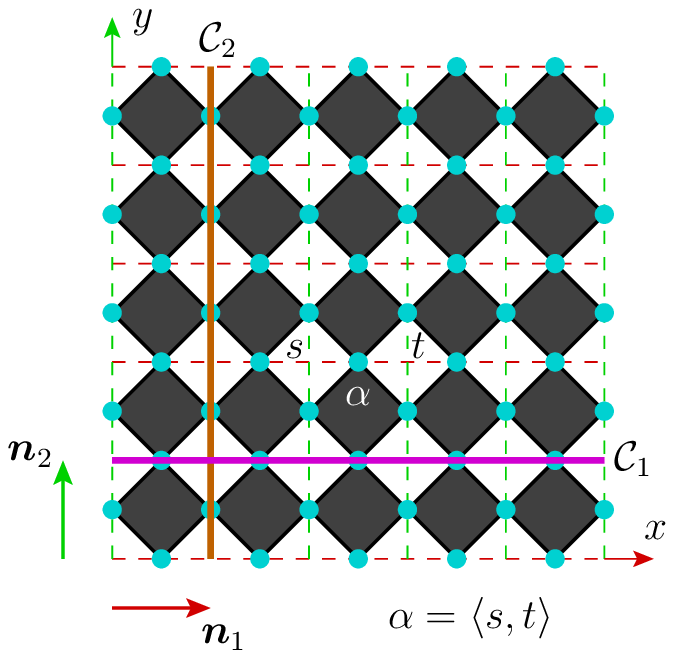}
  \caption{ (Color online)  Dots are located in the middle of the
    bonds of the original square lattice of Fig.~\ref{fig:lattice}
    (represented with dashed lines).  Bond operators of
    Eq.~(\ref{eq:bond_variables}) are defined on these bonds. White
    plaquettes (centered on vertices of the original lattice) are used
    to define $\widetilde{A}_s$ operators [see
    Eq.~(\ref{eq:As})],
    whereas $\widetilde{B}_p$ operators of Eq.~(\ref{eq:Bp}) are
    defined on gray plaquettes. The bond that is shared by two white
    plaquettes such as $s$ and $t$ is denoted by $\langle s, t
    \rangle$ or by a Greek letter such as $\alpha$. Contours
    $\mathcal{C}_1$ and $\mathcal{C}_2$ are used in
    Eq.~(\ref{eq:C1C2}) to define $\mathbb{Z}_2$ conserved quantities
    for a system with periodic boundary conditions.  }
  \label{fig:lattice2}
\end{figure}
%

As we have already seen in Sec.~\ref{sec:sub:TFIM} when describing the
spectrum of $H_\mathrm{I}$, the energy cost of any multi-magnon state
is given by the number of antiferromagnetic bonds of the state's spin
configuration. It is thus natural, though not mandatory, to introduce effective
spin variables living on the bonds (see Fig.~\ref{fig:lattice2}) as follows~:
%
\begin{equation}
  \label{eq:bond_variables}
  \widetilde{\sigma}_\beta^z
  = \widetilde{\sigma}_{\langle i, j \rangle}^z = \sigma_i^z \sigma_j^z.
\end{equation}
%
To make notations light, we denote bonds $\langle i, j \rangle$
between two sites $i$ and $j$ with Greek letters, such as $\beta$ in
the above equation.  A value $+1$ or $-1$ of
$\widetilde{\sigma}_\beta^z$ is then associated to, respectively, a
ferromagnetic or antiferromagnetic configuration of bond
$\beta$. Within such a description, Hamiltonian (\ref{eq:hamI})
becomes a pure field term
%
\begin{equation}
  \label{eq:hamI_mapped}
  \widetilde{H}_\mathrm{I}
  = -\frac{1}{2} \sum_\beta \widetilde{\sigma}_\beta^z.
\end{equation}
%

We can now give another form of the field term of $H_\mathrm{TFIM}$.
Since $\sigma_i^x$ flips the spin at site $i$, it flips the four
bonds sharing site $i$. Let us denote by $s(i)$ the set of these four
bonds (the notation $s$ referring to stars or vertices of the original
lattice).  Furthermore, we introduce
%
\begin{equation}
  \label{eq:As}
  \widetilde{A}_s = \prod_{\beta\in s(i)} \widetilde{\sigma}_\beta^x,
\end{equation}
%
which is the product of the four operators flipping bonds connected to site $i$. One can
alternatively say that the $ \widetilde{A}_s$ are defined on the white plaquettes
of Fig.~\ref{fig:lattice2}.  Then, the TFIM Hamiltonian reads
%
\begin{equation}
  \label{eq:hamTFIM_mapped}
  \widetilde{H}_\mathrm{TFIM}
  = -\frac{1}{2} \sum_\beta \widetilde{\sigma}_\beta^z
  - h \sum_s \widetilde{A}_s.
\end{equation}
%

In the same vein, we can rewrite the XX part of the XXZ Hamiltonian
(\ref{eq:hamXXZ_1}). From expression (\ref{eq:hamXXZ_3}), it is clear
that this XX term flips two adjacent spins, if these two spins are in
a ferromagnetic configuration while it annihilates antiferromagnetic bonds.
We recall that in Eq.~(\ref{eq:hamXXZ_3}), the sublattice rotation
was already used, so that a ferromagnetic configuration corresponds to
an original antiferromagnetic configuration in
Eqs.~(\ref{eq:hamXXZ_1}) and (\ref{eq:hamXXZ_2}). As a consequence, the XX
term flips the six bonds around a ferromagnetic bond, and one can
write
%
\begin{equation}
  \label{eq:hamXXZ_mapped}
  \widetilde{H}_\mathrm{XXZ}
  = -\frac{1}{2} \sum_\beta \widetilde{\sigma}_\beta^z
  + \lambda \sum_{\langle s,t\rangle} \widetilde{A}_s \widetilde{A}_t
  \frac{1+\widetilde{\sigma}_\alpha^z}{2},
\end{equation}
%
where the second sum runs over nearest-neighbor white plaquettes $s$
and $t$ that share a common bond $\alpha=\langle s, t\rangle$ (see
Fig.~\ref{fig:lattice2}), and involves projectors that annihilate an
antiferromagnetic configuration of such bonds.

\subsection{Symmetries, counting, and relation to toric code}
\label{sec:sub:symm_count}

Up to now, we have not yet discussed one fundamental aspect of the bond
description, namely, the counting of states. Indeed, if we assume that
the original lattice contains $N$ spins half $\boldsymbol{\sigma}$,
the mapping brings us to a description with $2N$ spins half
$\widetilde{\boldsymbol{\sigma}}$ since the number of bonds is twice
the number of sites (assuming periodic boundary conditions). Thus, it
seems that Hilbert spaces of the $\widetilde{H}$ Hamiltonians
(\ref{eq:hamI_mapped}), (\ref{eq:hamTFIM_mapped}), and
(\ref{eq:hamXXZ_mapped}) are too large, and involve ``unphysical"
states.

However, looking at these Hamiltonians a bit closer, it is clear that
the following operators are conserved
%
\begin{equation}
  \label{eq:Bp}
  \widetilde{B}_p = \prod_{\beta\in p} \widetilde{\sigma}_\beta^z.
\end{equation}
%
In this definition, the product is performed over all bonds belonging to $p$, which
can be any of the gray plaquettes shown in Fig.~\ref{fig:lattice2}
(they are also plaquettes in the original lattice shown in
Fig.~\ref{fig:lattice}). Among the $N$ $\mathbb{Z}_2$ operators that
can be defined this way, only $N-1$ can be set independently
to $\pm 1$, because of the constraint $\prod_p \widetilde{B}_p=1$ (with periodic boundary conditions).

With such notations, Hamiltonian (\ref{eq:hamTFIM_mapped}) is nothing
but the toric code Hamiltonian\cite{Kitaev03} in a magnetic field (we keep
using the $\widetilde{\phantom{a}}$ notations)
%
\begin{equation}
  \label{eq:hamTC}
  \widetilde{H}_\mathrm{TC} = -J_\mathrm{s} \sum_s \widetilde{A}_s
  -J_\mathrm{p} \sum_p \widetilde{B}_p
  -h_z \sum_\beta \widetilde{\sigma}_\beta^z ,
\end{equation}
%
with $J_\mathrm{s}=h$, $J_\mathrm{p}=0$ and $h_z=J=1/2$. Although the
original toric code (with nonvanishing $J_s$ and $J_p$) in a field has already been the subject of some works (see
Refs.~\onlinecite{Trebst07}, and \onlinecite{Hamma08,Tupitsyn08,Vidal09_1}), we are not
aware of any study of this precise model which has also been recently obtained in a related framework \cite{Cobanera10}. For this reason, we will
discuss some of its properties in Sec.~\ref{sec:TC}.

For periodic boundary conditions, as in the ``bare" toric code \cite{Kitaev03} (see also
Ref.~\onlinecite{Vidal08_2} for a pedagogical description), $\widetilde{B_p}$'s are not the only conserved operators when a
magnetic field in the $z$ direction is present.  Using contours $\mathcal{C}_1$ and $\mathcal{C}_2$
depicted in Fig.~\ref{fig:lattice2}, one can indeed define the following operators
%
\begin{equation}
  \label{eq:C1C2}
  \widetilde{C}_1 = \prod_{\beta\in\mathcal{C}_1} \widetilde{\sigma}_\beta^z,
  \qquad\mbox{and}\qquad 
  \widetilde{C}_2 = \prod_{\beta\in\mathcal{C}_2} \widetilde{\sigma}_\beta^z.
\end{equation}
%
These are also $\mathbb{Z}_2$ conserved quantities, which can be set
to $\pm 1$ independently of the values of the $\widetilde{B}_p$'s.

All in all, we have $(N-1)+2=N+1$ conserved and independent
$\mathbb{Z}_2$ quantities. Furthermore, to recover the ``physical"
subspace and the physics of the TFIM or the Heisenberg model, one should set all of them to $+1$. This reduces the
Hilbert space dimension from $2^{2N}$ to $2^{2N-(N+1)}=2^{N-1}$, that
is in fact less than the original Hilbert space dimension. However, the reason
for this obvious~: our description in terms of bond
variables is done with respect to one of the two degenerate
ferromagnetic ground states. As a consequence, such a description can
only be valid in the broken phase.

\section{Perturbative Continuous Unitary Transformations}
\label{sec:PCUT}

The aim of this section is to equip the reader with the basic knowledge about
PCUTs necessary for the understanding of the next section. For concreteness,  
we work out an example in detail, rather than focusing on a general framework.

\subsection{Basic ideas of continuous unitary transformations}
\label{sec:sub:basic_idea_CUT}

The CUTs method as known in the condensed-matter
theory community originates from the work of Wegner \cite{Glazek93,Glazek94,Wegner94}. A
pedagogical introduction to this technique can be found in Refs.~\onlinecite{Dusuel04_2} and \onlinecite{Dusuel05_2}.
The aim of this technique is to diagonalize or, more modestly, to block diagonalize a given
Hamiltonian $H$ thanks to a unitary transformation. The latter is not performed
in a single step but rather in a continuous way (whence the name of the
method) as
%
\begin{equation}
    H(l) = U^\dagger(l) H U(l),
    \label{eq:Hl}
\end{equation}
%
where $l$ is a running parameter such that $H=H(l=0)$ and
$H_\mathrm{eff}=H(l=\infty)$ is an effective (block-) diagonal Hamiltonian.
This equation can be cast into a differential (flow) commutator equation \cite{Wegner94}
%
\begin{equation}
    \label{eq:dlH}
    \partial_l H(l) = [\eta(l), H(l)],
\end{equation}
%
where $\eta(l)=\partial_l U^\dagger(l) U(l)$ is the anti-Hermitian generator
associated to the unitary transformation $U(l)$.

\subsection{Quasi-particle conserving generator}
\label{sec:sub:QPC_generator}

The next task is to find the appropriate generator which, from the local
knowledge of $H(l)$, leads to the desired form of the effective Hamiltonian.
We shall only discuss the QP conserving generator \cite{Knetter00_1}
that will be used in the sequel. We furthermore focus on a specific example
for which the Hamiltonian can be written 
%
\begin{equation}
    \label{eq:H024}
    H = Q + \sum_{n=-n_\mathrm{max}}^{+n_\mathrm{max}} T_n.
\end{equation}
%
For concreteness, in the following we consider the case where $n\in\{0,\pm 2, \pm 4\}$ which is relevant for the TFIM.
In this equation, $Q$ is the Hermitian operator which counts the number of
QPs (so its spectrum is contained in $\mathbb{N}$), and the $T_n$'s are
operators that change the QP number by the amount $n$, so that $[Q,
T_n] = n T_n$. The hermiticity of the Hamiltonian requires that $T_n^\dagger =
T_{-n}$. The QP conserving generator is designed to bring the
Hamiltonian to an effective form that conserves the number of QPs~:
$[Q,H_\mathrm{eff}]=0$.  Said differently, under the CUTs, all terms $T_n$ will be flowing
(but not $Q$ which is isolated from all other terms), thus becoming
$T_n(l)$, and one wishes to reach a situation where $T_n(l=\infty)=0$ for all
$n\neq 0$. This can be achieved \cite{Knetter00_1, Dusuel04_2, Dusuel05_2} by
choosing
%
\begin{equation}
    \label{eq:eta024}
    \eta(l) = T_{+2}(l) - T_{-2}(l) + T_{+4}(l) - T_{-4}(l).
\end{equation}
%
With this choice of generator, the flow Eq.~(\ref{eq:dlH}) can be written
%
\begin{eqnarray}
    \label{eq:flow024}
    \partial_l T_0(l) &=&
    2 \big[T_{+2}(l), T_{-2}(l)\big] + 2 \big[T_{+4}(l), T_{-4}(l)\big], \nonumber\\
    \partial_l T_{+2}(l) &=& -2 T_{+2}(l)
    + \big[T_{+2}(l), T_0(l)\big] \nonumber \\
    && +2 \big[T_{+4}(l), T_{-2}(l)\big], \nonumber\\
    \partial_l T_{+4}(l) &=& -4 T_{+4}(l)
    + \big[T_{+4}(l), T_0(l)\big].
\end{eqnarray}
%
We have not written the flow equations for $T_{-2}(l)$ and $T_{-4}(l)$ since
the Hamiltonian remains Hermitian under a unitary transformation. Let us
emphasize that no new term appears during the flow, thanks to the
choice of the generator ({\it e.~g.}, no term creating six particles
appears since $[T_{+2}(l), T_{+4}(l)] + [T_{+4}(l), T_{+2}(l)]=0$). Linear terms in the
right-hand side of Eq.~(\ref{eq:flow024}) ensure that \mbox{$T_{n \neq 0}(l=\infty)=0$} so that $[H_\mathrm{eff},Q]=0$.

\subsection{Perturbative commutator expansion of the flow equation}
\label{sec:sub:pert_flow_eq}

It remains to solve these flow equations which is still a challenging task since the $T_n(l)$ terms contain
an infinite number of operators. 
The easiest way of doing so is to perform a perturbative expansion of the flow equations, assuming that all $T_n$ terms in $H$ [see Eq.~(\ref{eq:H024})] are small, and of the same order of
magnitude. 
Such an expansion was first performed by Stein \cite{Stein97} for a Hamiltonian without $T_{\pm 4}$ terms, and extended to Hamiltonians with terms changing
$Q$ by any amount by Knetter and Uhrig \cite{Knetter00_1} who also provided a description in terms of QP. However the formalism
used in these papers was not the same as the one presented here, and, in
particular, the emphasis was not on getting a commutator expansion, which is our
goal in the following.

To this end, the expansion for $T_n(l)$ is simply written
%
\begin{equation}
    \label{eq:exp_Hn}
    T_n(l) = \sum_{i=1}^\infty T_n^{(i)}(l),
\end{equation}
%
where the superscript $(i)$ is the order, in perturbation, of $T_n^{(i)}(l)$.
The flow equations can then be expanded as

%
\begin{eqnarray}
    \label{eq:flow024_pert}
    \partial_l T_0^{(i)}(l) &=&
    \sum_{j=1}^{i-1}\left(
    2 \left[T_{+2}^{(j)}(l), T_{-2}^{(i-j)}(l)\right]\right.\nonumber\\
    && \hspace{0.5cm}
    \left. + 2 \left[T_{+4}^{(j)}(l), T_{-4}^{(i-j)}(l)\right]\right),\nonumber\\
    \partial_l T_{+2}^{(i)}(l) &=& -2 T_{+2}^{(i)}(l) + \sum_{j=1}^{i-1}\left(
    \left[T_{+2}^{(j)}(l), T_0^{(i-j)}(l)\right]\right.\nonumber\\
    && \hspace{2.2cm}
    \left. + 2 \left[T_{+4}^{(j)}(l), T_{-2}^{(i-j)}(l)\right]\right), \nonumber\\
    \partial_l T_{+4}^{(i)}(l) &=& -4 T_{+4}^{(i)}(l)
    + \sum_{j=1}^{i-1} \left[T_{+4}^{(j)}(l), T_0^{(i-j)}(l)\right].
\end{eqnarray}
%
These have to be solved with the initial conditions $T_n^{(i)}(l=0)=\delta_{i,1}T_n$ and then one can take the limit $l\to\infty$ to
obtain $H_\mathrm{eff}$. For the Hamiltonian considered up to
now, one obtains, at order 3, the following commutator expansion
%
\begin{eqnarray}
    \label{eq:Heff024_order3}
    H_\mathrm{eff} &=& Q + T_0 + \frac{1}{2}[T_{+2}, T_{-2}]
    + \frac{1}{4}[T_{+4}, T_{-4}] \\
    && + \frac{1}{8}\Big(\big[[T_{+2},T_0],T_{-2}\big]
    +\big[T_{+2},[T_0,T_{-2}]\big]\Big)\nonumber\\
    && + \frac{1}{8}\Big(\big[[T_{+4},T_{-2}],T_{-2}\big]
    +\big[T_{+2},[T_{+2},T_{-4}]\big]\Big)\nonumber\\
    && + \frac{1}{32}\Big(\big[[T_{+4},T_0],T_{-4}\big]
    +\big[T_{+4},[T_0,T_{-4}]\big]\Big). \nonumber
\end{eqnarray}
%

We have computed such expansions for Hamiltonians of the form given in
Eq.~(\ref{eq:H024}) to various orders given in Table \ref{tab:orders_coeff}.
In this table, we also give the total number of nonzero coefficients one
obtains once the commutators in the effective Hamiltonian have been expanded in
polynomials in $T$ operators. Let us note that for the Hamiltonian we have been
considering up to now, with $n\in\{-4,-2,0,2,4\}$, the expansion can be obtained
from the one of $n_\mathrm{max}=2$ ({\it i.~e.}, $n\in\{-2,-1,0,1,2\}$) by a
proper rescaling of the corresponding coefficients.
%
%
\begin{table}[b]
    \centering
    \begin{tabular}{||c|c|c||}
        \hline
        \hline
        $n_\mathrm{max}$ & Order & Number of coefficients \\
        \hline
        \hline
        1 & 18 & 67214380 \\
        \hline
        2 & 14 & 569842124 \\
        \hline
        3 & 12 & 924457284 \\
        \hline
        4 & 10 & 189956506 \\
        \hline
        \hline
        \end{tabular}
    \caption{Maximum order at which we have derived the effective Hamiltonian, for
    various values of $n_\mathrm{max}$ [see Eq.~(\ref{eq:H024}) for definitions], as well as
    the total number of non-zero coefficients needed to express the effective
    Hamiltonian as a polynomial in $T_n$'s.
   }
    \label{tab:orders_coeff}
\end{table}
%
%

Let us also mention that all coefficients are given by rational numbers and are valid for arbitrary system size, including the thermodynamical limit.
The very last (but not least) step in the whole computation is to apply this effective
Hamiltonian to states with fixed number of QPs and to diagonalize it.

To determine the low-energy QP properties of the TFIM and the
XXZ Hamiltonian around the Ising limit, we shall use this PCUTs approach. 
Indeed, in this limit, both systems can be written in the form (\ref{eq:H024}) by defining 
the operator
%
%
\begin{equation}
  \label{eq:Q}
  Q = \sum_\beta \frac{1 - \widetilde{\sigma}_\beta^z}{2}
  = N+\widetilde{H}_\mathrm{I},
\end{equation}
%
%
which counts the number of antiferromagnetic bonds.
The operators $T_n$ are then proportional to the perturbation ($h$ in the TFIM
and $\lambda$ in the XXZ model). The index $n$ denotes the change in the
number of antiferromagnetic bonds $q$. For the TFIM one has $n\in\{0, \pm 2,
\pm 4\}$ and for the XXZ model one gets $n\in\{0, \pm 2, \pm 4, \pm 6\}$. The
larger number of $T_n$ operators for the XXZ model results in a larger effort
since more processes have to be taken into account for a given perturbation
order. For instance, for the 2QP sector (containing bound states), we reached
order 12 for the TFIM and order 8 for the XXZ model.

Here, we use the transformed Hamiltonians (\ref{eq:hamTFIM_mapped}) and (\ref{eq:hamXXZ_mapped}) defined with bond variables, Eq.~(\ref{eq:bond_variables}), but PCUTs can be (and for actual computer implementations are) applied
directly to Eqs.~(\ref{eq:hamTFIM}) and (\ref{eq:hamXXZ_3}). In order to avoid
possible confusion, we shall use the notation 0QP, 1QP, and 2QP when referring to
$0$, $1$, and $2$ magnons, and 0qp, 1qp, \ldots, 8qp when referring to $0$, $1$,
\ldots,$8$ antiferromagnetic bonds. 
The ground state of the effective Hamiltonian is the state without antiferromagnetic bonds
($q=0$). The one-magnon excitations have $q=4$ and bound states of two
nearest-neighbor magnons have $q=6$ antiferromagnetic bonds. Things become much
more complicated for configurations with more antiferromagnetic bonds, and we
shall not study them here. 
Let us simply mention that $q=8$ corresponds, in the unperturbed limit, either to two unbound magnons
or to three- or four-magnon bound states. 
However, at finite coupling, we cannot exclude that there also exist two-magnon bound
states (different from those discussed above) in this sector.

\section{Linked-cluster expansion and quantum finite-lattice method}
\label{sec:LCE_QFL}

\subsection{Linked-cluster expansion}
\label{sec:sub:LCE}

%
\begin{figure}[b]
  \includegraphics[width=\columnwidth  ]{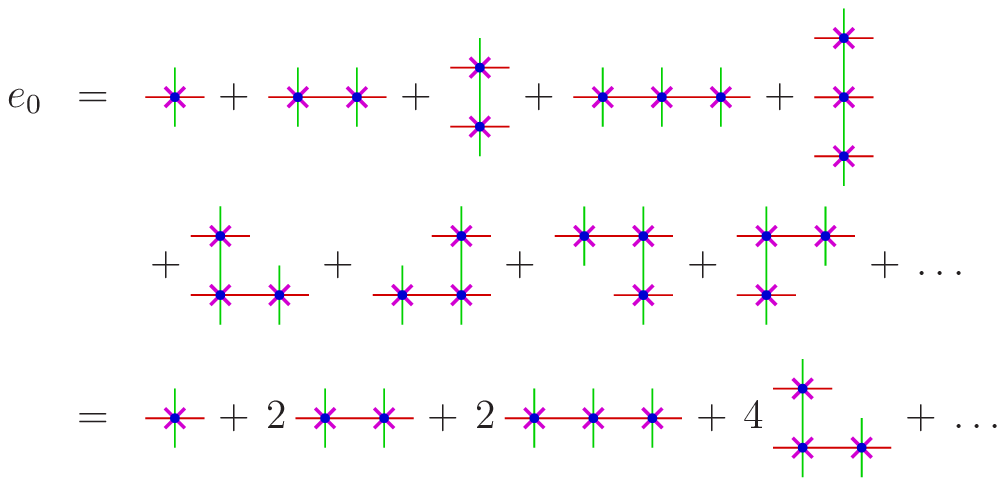}
  \caption{ (Color online)  Pictorial representation of the ground-state energy
  per site as a linked-cluster expansion. A cross on a site means that this site
  has been acted upon at least twice by a $T_n$ operator (two spin flips per
  site are necessary to preserve the ferromagnetic state). The second line
  follows from the symmetries of the lattice. All contributions relevant for a
  computation at order $3$ of $e_0$ are shown, contrarily to contributions of
  order $4$ (which are all gathered in ``$\ldots$").}
  \label{fig:graphology}
\end{figure}
%

Flow Eq.~(\ref{eq:flow024_pert}) show that $H_\mathrm{eff}$ can be
written as a perturbative commutator expansion to any order, as exemplified in
Eq.~(\ref{eq:Heff024_order3}) to order 3. This property has dramatic consequences when
one considers a Hamiltonian defined on a lattice with local $T_n$ operators.
By local, we mean that $T_n=\sum_i T_{n,i}$, where $i$ runs over the lattice
sites, with $T_{n,i}$ acting on a finite number of sites neighboring $i$.
Indeed, in such a situation, commutators $[T_{n,i},T_{p,j}]$ vanish as soon
as $i$ and $j$ are sufficiently far apart and are local operators. One
could try to implement the calculation of the commutators in a symbolic way but
it is usually much easier to apply the effective Hamiltonian (with expanded commutators) to
states. For the practical purpose of evaluating the action of one term in
$H_\mathrm{eff}$, one then only needs to apply $H_\mathrm{eff}$ to a finite-size
linked cluster of sites. The notion of linked is problem-dependent since it
depends on the extension of $T_{n,i}$ operators.

The appearance of a linked-cluster expansion for effective Hamiltonians written
as commutators is a long-established and well-known result (see {\it e.~g.},
Ref.~\onlinecite{Primas63}), although it seems to have escaped the attention it
deserves in more recent publications \cite{Gelfand00, Oitmaa06, Knetter03_1}.

Let us finally note that the PCUTs method is not the only available one to obtain such a perturbative
commutator expansion and one could use the Van Vleck formalism (see
{\it e.~g.}, Ref.~\onlinecite{Shavitt80}) as well. The main advantage of PCUTs,
 is that it does not generate terms creating any number of QPs, so that the bookkeeping is not too
difficult. One drawback is that one has to solve flow equations instead of
performing purely algebraic manipulations.

In the following, we will introduce the two commonly used implementations of 
the linked-cluster theorem which represent extreme cases. Either a calculation is performed
 on a large number of minimal clusters or a calculation is done on one very large 
cluster. Afterwards, a third implementation of the linked-cluster theorem is presented which 
is a compromise and which we use in the current study. 

\subsection{Graph-ology}
\label{sec:sub:graphology}

%
\begin{figure}[t]
  \includegraphics[width=\columnwidth  ]{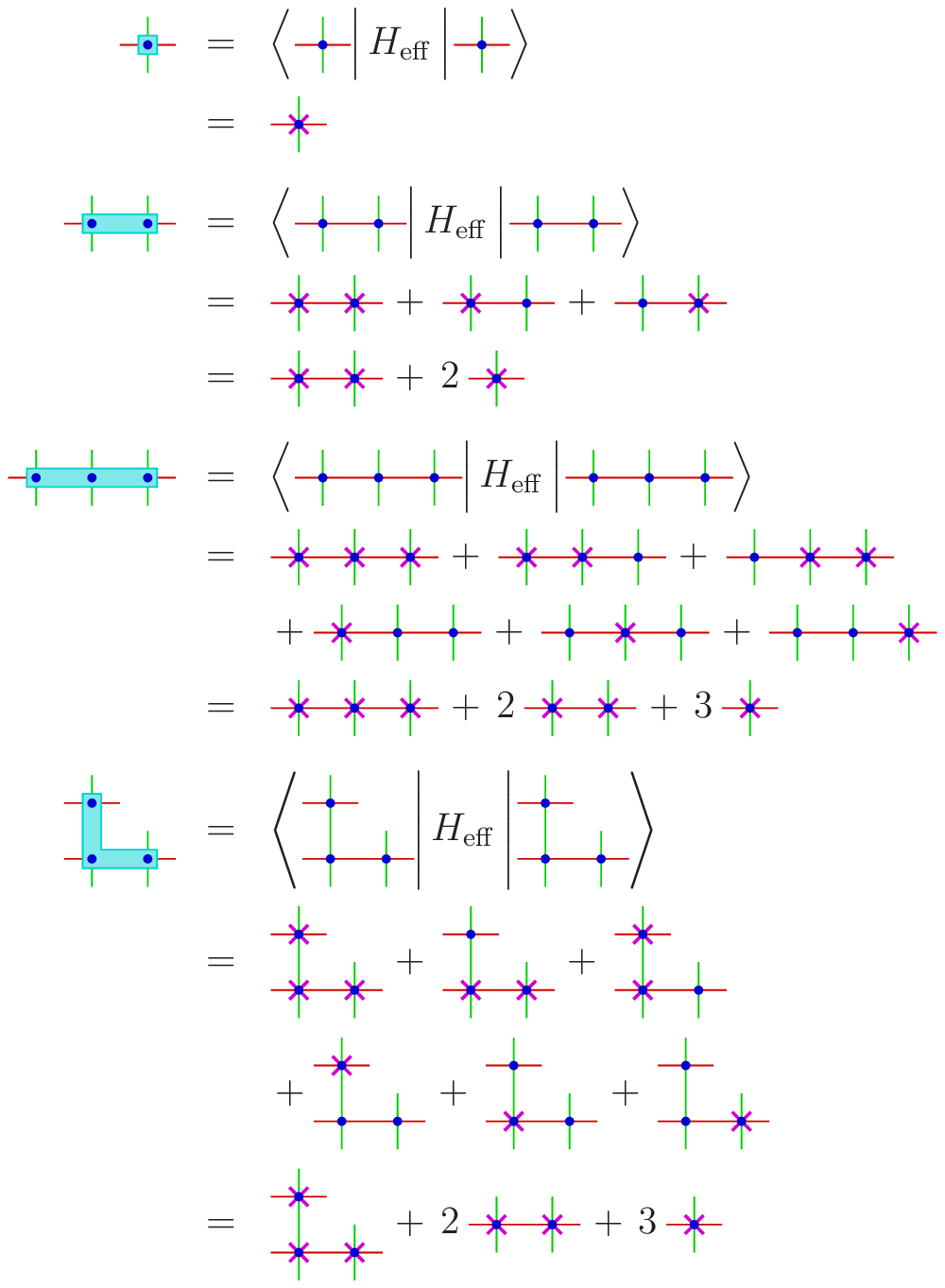}
  \caption{ (Color online)  Pictorial representation of the computation of
  matrix elements of $H_\mathrm{eff}$ for various clusters. Crosses have
  the same meaning as in Fig.~\ref{fig:graphology}. Use of symmetry is made to
  simplify equations and to compute only four graphs (see
  Fig.~\ref{fig:graphology}) but others can be computed as easily.}
  \label{fig:graphology_matrix_elements}
\end{figure}
%

With Sec.~\ref{sec:sub:LCE} in mind, it should now be clear that the aim is to
compute contributions of the effective Hamiltonian when the latter is allowed to
act on states defined on clusters of one site, two sites, etc.  To make things
more concrete, in what follows, we shall focus on the computation of the
ground-state energy per site $e_0$ of the TFIM at order $6$ in $h$.  As only
even powers of $h$ will appear in the expansion, we shall for simplicity only
refer to the order in $h^2$ in this section, so that we will say that we work at order $n$ when 
computing terms up to order $(h^2)^n$. Only the action on the ferromagnetic 
(0QP) state will be considered here so that we
do not draw any arrows to make things clearer. The relevant contributions from
different clusters are shown in Fig.~\ref{fig:graphology}. Let us mention that
the clusters shown in this figure do have contributions at order $4$ or more (in
which case at least one site is acted upon twice) but clusters not shown
(with four crosses or more) do not have contributions at order $3$ or less. To
obtain the last line of Fig.~\ref{fig:graphology}, the symmetries of the
lattice have been used. Note that the $T_n$ operators can act on at most three
sites (each pictogram has at most 3 crosses in Fig.~\ref{fig:graphology}) at order $3$ and in
the 0QP sector, because each $T_n$ operator flips the spin of the site on which
it acts, so one needs two $T_n$ operators per site to start from and end up with a
ferromagnetic (0QP) state.

\begin{figure}[t]
  \includegraphics[width=\columnwidth]{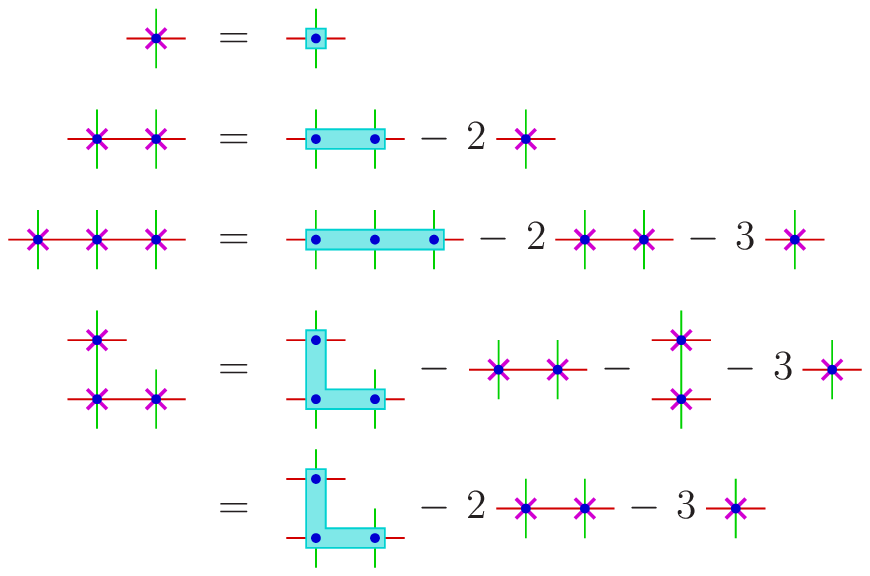}
  \caption{ (Color online)  Pictorial representation of the inversion 
  of the relations of Fig.~\ref{fig:graphology_matrix_elements}, which leads to
  subcluster subtraction. The last equality is due to symmetries.}
  \label{fig:graphology_subtraction}
\end{figure}

To compute the contributions shown in Fig.~\ref{fig:graphology}, it is not very
practical to make sure that all sites have been acted upon at least once by a $T_n$
operator. It is much easier in actual computations to compute all possible
outcomes of the action of $H_\mathrm{eff}$ on a given cluster and to subtract
contributions of subclusters, as illustrated in
Figs.~\ref{fig:graphology_matrix_elements} and \ref{fig:graphology_subtraction}.
Let us note that these subtractions are mandatory in the perturbation theory
used in Refs.~\onlinecite{Gelfand00} and \onlinecite{Oitmaa06} which is not based on an
expression of $H_\mathrm{eff}$ as a series in $T_n$ operators. Numerical
coefficients appearing in Fig.~\ref{fig:graphology_subtraction} are
simply the number of ways the subclusters (or their symmetric-related ones) can
be embedded in a given cluster.

%
%

Thus, we are naturally led to enumerate all possible
linked clusters (\textit{i.~e.}, also graphs, hence the name of this section) that
can be embedded in the square lattice, and apply the subtraction technique to
each of the cluster. Though this is the least memory-consuming way to proceed,
and though the computational effort required for the computation on one cluster
is rather small, it requires to perform a heavy and time-consuming combinatorial
work. One indeed has to enumerate all possible linked clusters, and then, for
each cluster, one needs to find all its linked subclusters. In order for this
technique to be as efficient as possible, one should also ensure that
topologically identical clusters are identified (such as the two three-site
clusters of Fig.~\ref{fig:graphology_subtraction}).  This again reduces
the computational effort but makes the combinatorial tasks even harder and more
time consuming.

A completely opposite way of performing the calculation is to reduce the
combinatorial complexity to zero by computing all contributions in one go,
thanks to a cluster with periodic boundary conditions, large enough so that it
can accommodate all clusters one is interested in, without any finite-size
effect. In this way, applying $H_\mathrm{eff}$ to a 0QP state defined on such a
cluster one recovers the same state, up to a multiplicative factor being equal
to $e_0$ times the number of sites of the cluster. However, as the
Hilbert-space size quickly grows with the number of sites, this makes all
computations greedy in memory, but also in time (because the number of
intermediate states generated when applying $H_\mathrm{eff}$ can become huge).

\subsection{Quantum finite-lattice method}

%
\begin{figure}[t]
  \includegraphics[width=\columnwidth  ]{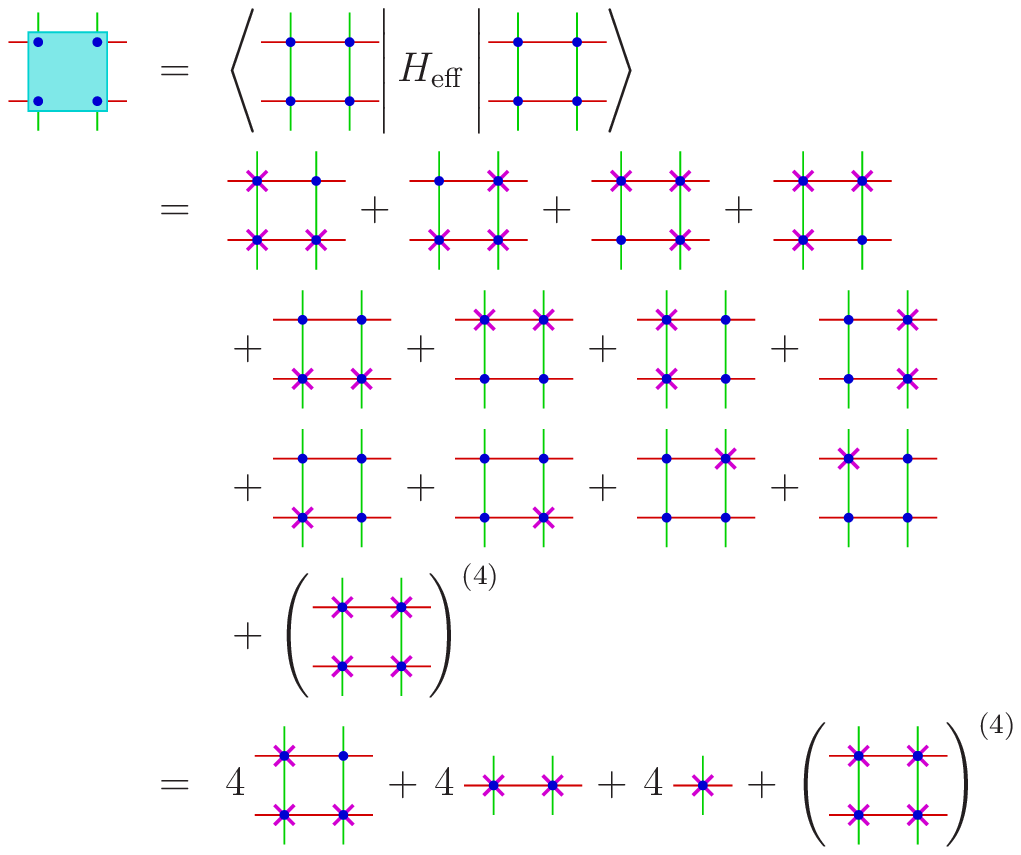}
  \caption{ (Color online)  Fourth (and last) matrix element from the $2\times 2$
  rectangular cluster to be considered for a finite-lattice computation at order $3$
  (apart from the first three of Fig.~\ref{fig:graphology_matrix_elements}).
  The term in parentheses involves the action on four sites and only appears at
  order $4$ and beyond (whence the exponent of the parenthesis).}
  \label{fig:FL_matrix_elements}
\end{figure}
%

We shall now see how one can work half-way between these two extreme cases.
The idea is to generalize Enting's finite-lattice method (see
Refs.~\onlinecite{deNeef77,Enting96,Oitmaa06}, and references therein)
from the statistical physics setting (where it is used to compute the free
energy) to the realm of quantum physics. Essentially, the idea is to use
rectangular clusters only and to perform appropriate subtractions. The main
advantage of using rectangular clusters is that computing the number of
embeddings of rectangular subclusters is trivial. Compared to graph-ology, one
gains enormously from the combinatorial side, but of course one has to pay a
price~: the clusters one uses are larger than those of graph-ology. Enting's
finite-lattice method is most useful in low dimensions (namely, two dimensions in
statistical mechanics) where the graph sizes do not grow too fast with the order
in perturbation but with modifications it can also be efficiently applied to
higher dimensions (see Ref.~\onlinecite{Arisue03} for an application in three
dimensions).

Let us first discuss this method in the 0QP sector, where it is almost the same, in the context we are working in, as the original method of statistical mechanics, because the ground state is not infinitely degenerate. In the latter
case, our method would provide an effective Hamiltonian in the low-energy
subspace which is a major difference compared to a single number (the
ground-state energy or the free energy in statistical mechanics). The first
three equations of Fig.~\ref{fig:graphology_matrix_elements} remain valid since they
involve a rectangular cluster but the last equation is now replaced by the new
matrix element of Fig.~\ref{fig:FL_matrix_elements}.

%
\begin{figure}[t]
  \includegraphics[width=\columnwidth ]{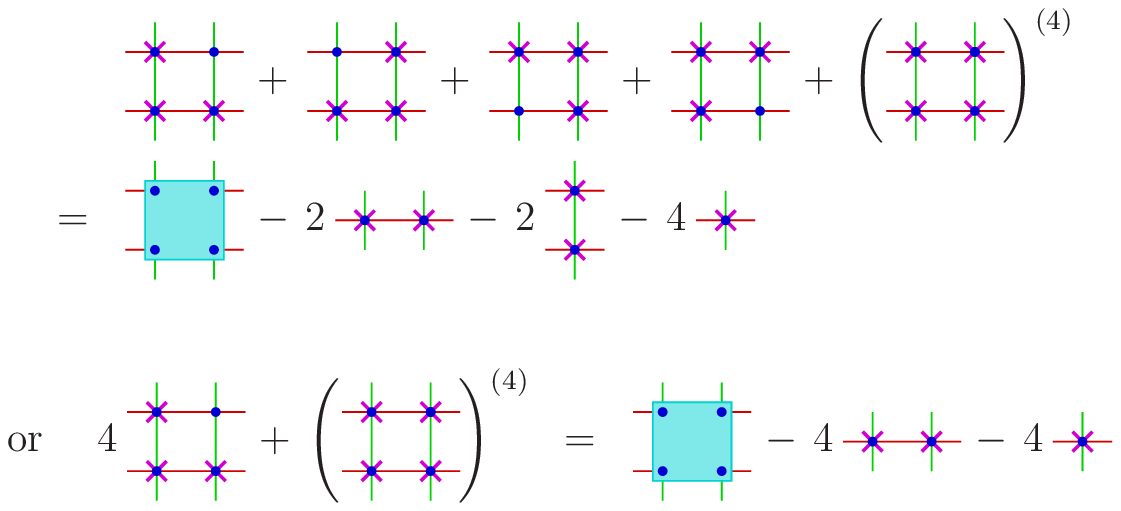}
  \caption{ (Color online)  Inversion of the equality of
  Fig.~\ref{fig:FL_matrix_elements}, with the help of the first three lines
  of Fig.~\ref{fig:graphology_subtraction}, showing how subcluster subtraction
  appears.
  The second equality makes use of symmetries, contrarily to the first line.}
  \label{fig:FL_subtraction}
\end{figure}
%

This can again be inverted to yield subcluster subtraction and is shown in
Fig.~\ref{fig:FL_subtraction}. \textit{From this, one understands that
a rectangular cluster yields, after subtraction, the sum of all contributions of
the linked subclusters that cannot be embedded in any rectangular subcluster (of
the considered cluster)}. If the cluster has width or height $1$, the sum
actually consists in a single term. From these considerations, one can deduce a simple
formula. Let us call $D_{s_x,s_y}$ the expectation value of $H_\mathrm{eff}$, computed on a
rectangular cluster of size $(s_x,s_y)$, so that the first three equations of
Fig.~\ref{fig:graphology_matrix_elements} and the one of Fig.~\ref{fig:FL_matrix_elements}
correspond to $D_{1,1}$, $D_{2,1}$, $D_{3,1}$, and $D_{2,2}$. 
Let us furthermore
define (recursively)
%
\begin{equation}
  \label{eq:FL_sbt}
  \widetilde{D}_{s_x,s_y} = D_{s_x,s_y} -
  {\sum_{t_x,t_y}}'(s_x-t_x+1)(s_y-t_y+1)\widetilde{D}_{t_x,t_y},
\end{equation}
%
where $\sum'$ means that the sum is restricted to the set of strict
subclusters
of sizes $(t_x,t_y)$ of the cluster of size $(s_x,s_y)$, namely, satisfying
$1\leqslant t_x\leqslant s_x$ and $1\leqslant t_y\leqslant s_y$ with
$(t_x,t_y)\neq (s_x,s_y)$.
As an illustration, the first three lines of
Figs.~\ref{fig:graphology_subtraction} and \ref{fig:FL_subtraction} give $\widetilde{D}_{1,1}$,
$\widetilde{D}_{2,1}$,
$\widetilde{D}_{3,1}$, and $\widetilde{D}_{2,2}$.
 Then the ground-state energy
can be expressed as a sum of $\widetilde{D}_{s_x,s_y}$ contributions, for a
range of $s_x$ and $s_y$ that is problem dependent (because the $T_n$ operators
can have different spatial extensions, as is the case for the TFIM and the
Heisenberg model). In the case of the TFIM we have been focusing on up to now, one has
%
\begin{equation}
  e_0= e_0^{(0)}+\sum_n e_0^{(n)},
\end{equation}
%
with
%
\begin{equation}
  \label{eq:FL_e0}
  e_0^{(n\geqslant 1)}
  = \sum_{1\leqslant s_x+s_y-1\leqslant n} \widetilde{D}_{s_x,s_y},
\end{equation}
%
and $e_0^{(0)}$ is the constant shift of order $0$ in the Hamiltonian.
Contrary to the TFIM where $T_n$ operators act on one site, the $T_n$'s for the Heisenberg model act on two sites
so that the sum for $e_0^{(n)}$ would be restricted to \mbox{$1\leqslant s_x+s_y-1\leqslant 2n$}, where $n$ is the order in $\lambda^2$.
%
\begin{figure}[t]
  \includegraphics[width=\columnwidth ]{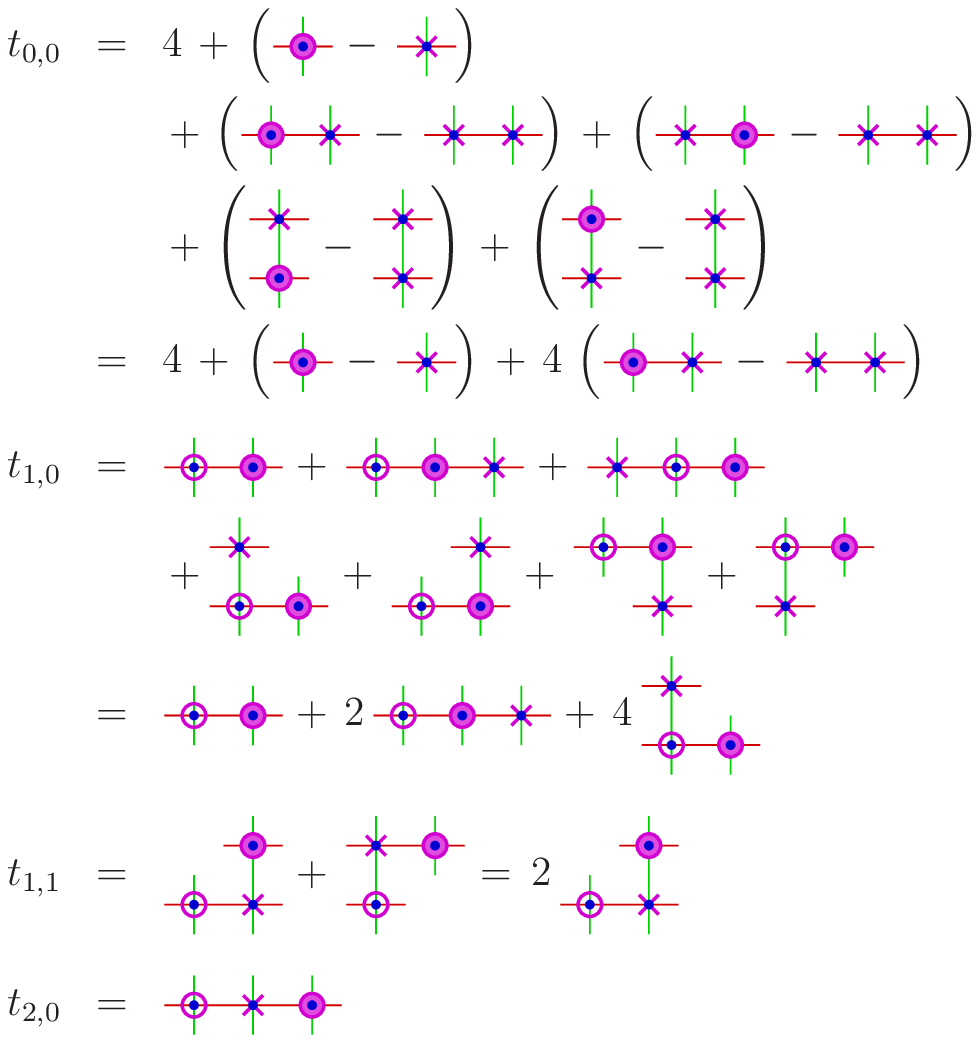}
  \caption{ (Color online)  Contributions to hopping amplitude at order 2, as a
  linked-cluster expansion. Crosses have the same meaning as in
  Fig.~\ref{fig:graphology}, and are acted upon at least twice by $T_n$
  operators. The empty (full) circle represents the initial (final) position of
  the particle. Sites with circles are acted upon at least once by $T_n$
  operators. When the ``particle" does not move, \textit{i.~e.}, for $t_{0,0}$, the
  particle's site (only represented with a full circle) is acted upon at least
  twice by $T_n$ operators. In this case, as one is only interested in the
  energy of a 1QP state with respect to the 0QP ground state, one must subtract
  the 0QP amplitude (Refs.~\onlinecite{Gelfand00,Oitmaa06} and \onlinecite{Knetter03_1}). Note that for
  $t_{0,0}$, one should also not forget the action of the counting operator $Q$,
  which gives a zeroth-order contribution equal to $4$.} 
  \label{fig:hoppings}
\end{figure}
%

The very same idea can be applied for any QP sector, provided appropriate
subtractions of contributions from sectors with lower number of QPs are
performed, as in standard linked-cluster expansions
\cite{Gelfand00, Oitmaa06, Knetter03_1}. In the same spirit as before, let us
give a pictorial explanation based on the TFIM, at order 2, and for the 1QP
sector (the 2QP sector can be worked out in a similar way). The aim is to
compute hopping amplitudes, as those given in Appendix
\ref{app:sub:TFIM_hopping1}.
As can be seen in Fig.~\ref{fig:hoppings}, one difference
compared to the 0QP sector is that depending on which term of the Hamiltonian
one computes, the particle can hop to different final positions. In the figure,
we have again not represented any arrow to make things light, but it should be
clear that the reference state is a ferromagnetic state, with one spin flip at
the QP's position. We have only considered hoppings $t_{i,j}$ of $i$ sites in
$x$ direction and $j$ sites in $y$ direction, for which $i \geqslant j \geqslant 0$,
other processes can be found thanks to symmetries of the square lattice and
the hermiticity of the Hamiltonian. The hopping amplitude $t_{0,0}$, which
should rather be called a chemical potential, is a bit peculiar. First, the
counting operator $Q$ appearing in $H_\mathrm{eff}$ gives a contribution of $4$.
Second, as one is only interested in the excitation energy above the ground
state, the 0QP contributions of the clusters have to be subtracted.

\begin{figure}[t]
  \includegraphics[width=\columnwidth ]{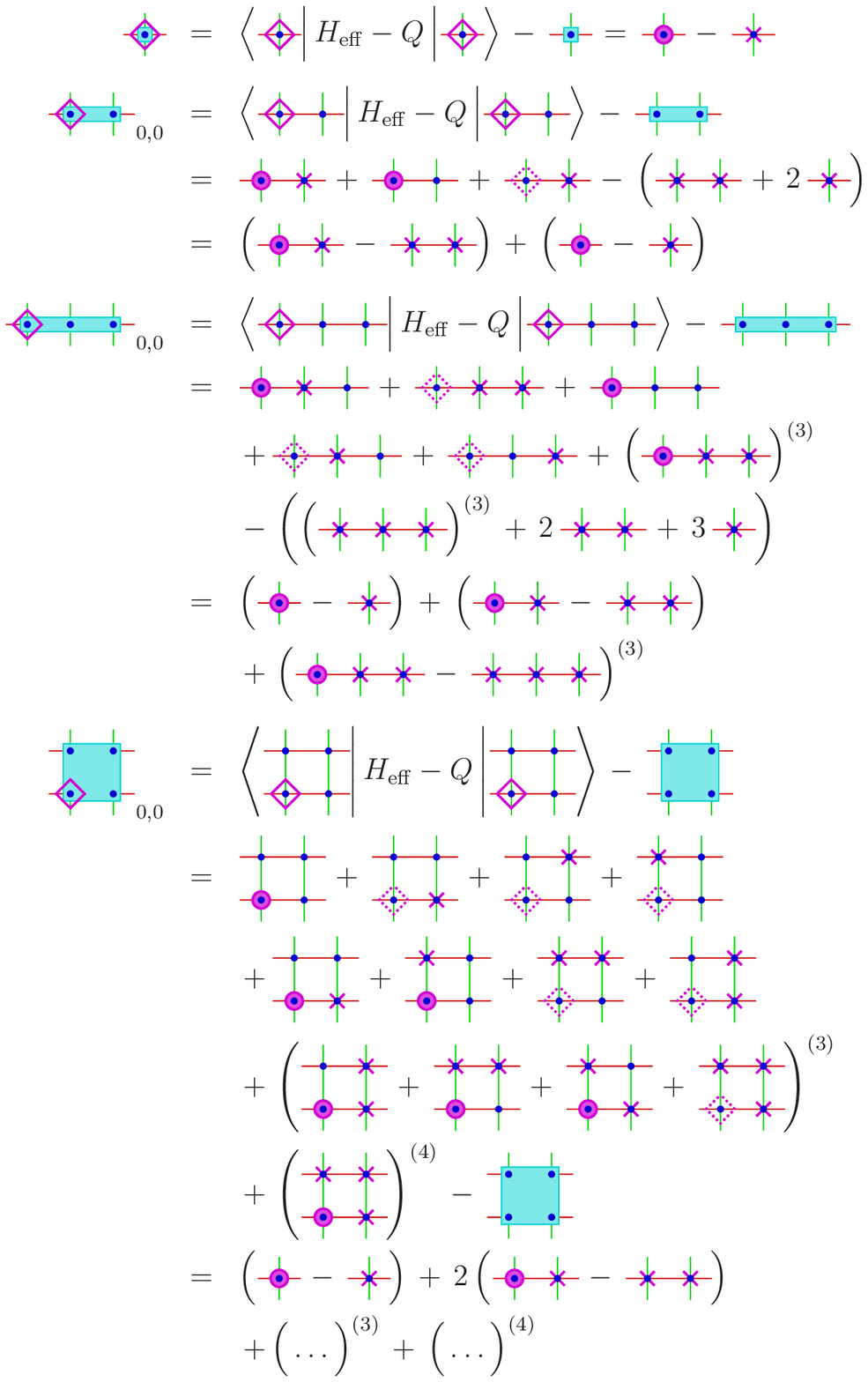}
  \caption{ (Color online)  Matrix elements of $H_\mathrm{eff}$ between
  identical initial and final states that are useful for the computation of
  $t_{0,0}$ at order 2. Diamonds with solid lines represent the particle position in the considered state.
  Contributions arising from an action of  $H_\mathrm{eff}-Q$ which does not involve the
  particle site contain a dotted diamond that keeps trace of this site.
 Other symbols have the same meaning as in previous figures. Terms in parenthesis are higher-order
  terms, for which the exponent gives the lowest order they start to contribute.
  For the last contribution, intermediate steps have not been shown, and the
  higher-order terms are not kept until the end, to keep things as light as possible.}
  \label{fig:hoppings_matrix_elements_00}
\end{figure}

\begin{figure}[t]
  \includegraphics[width=\columnwidth ]{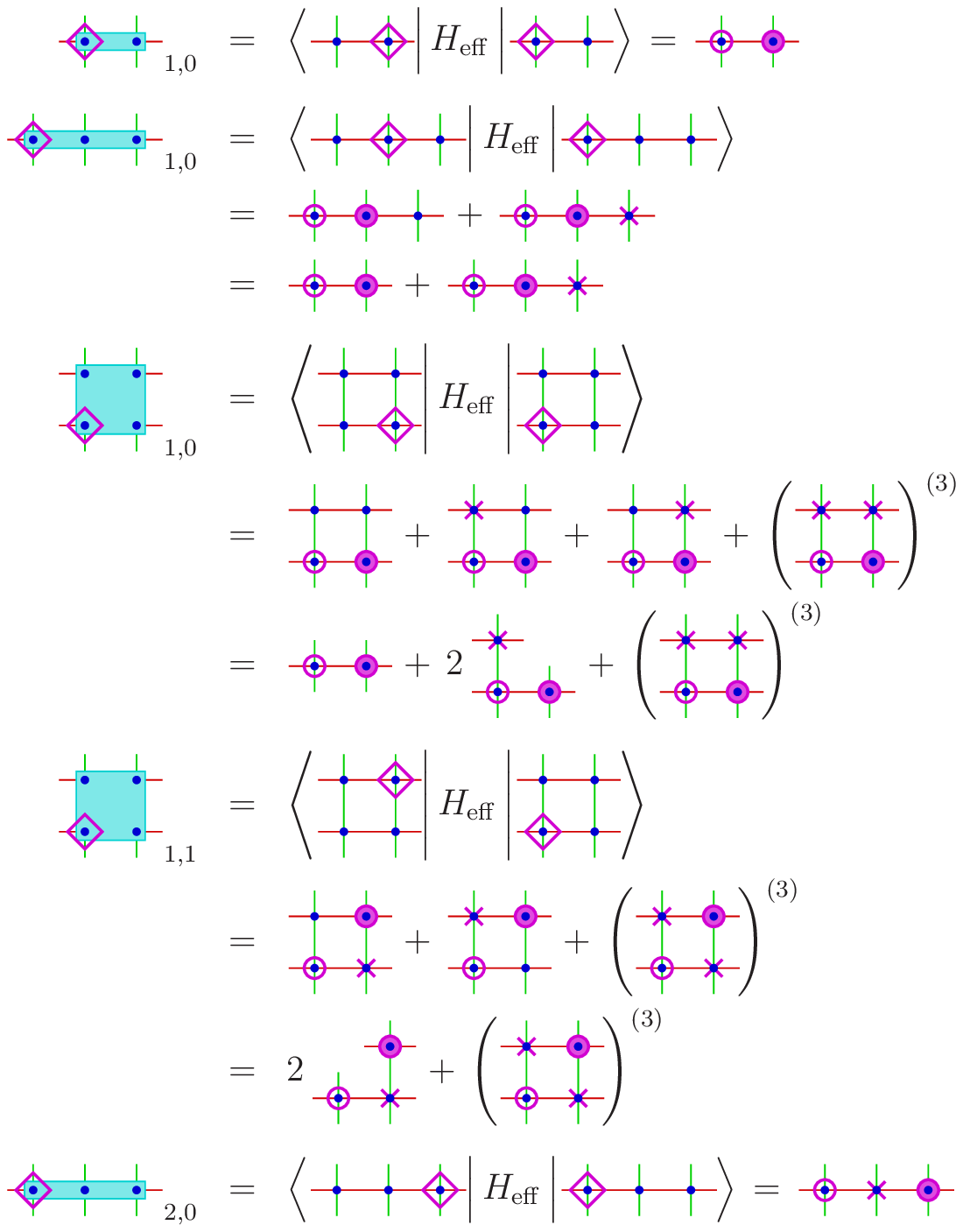}
  \caption{ (Color online)  Matrix elements of $H_\mathrm{eff}$ between 
  different initial and final states that are useful for the computation of
  $t_{1,0}$, $t_{2,0}$, and $t_{1,1}$ at order 2. Conventions are the same as in 
  Fig.~\ref{fig:hoppings_matrix_elements_00}.}
  \label{fig:hoppings_matrix_elements_n00}
\end{figure}

\begin{figure}[t]
  \includegraphics[width=\columnwidth]{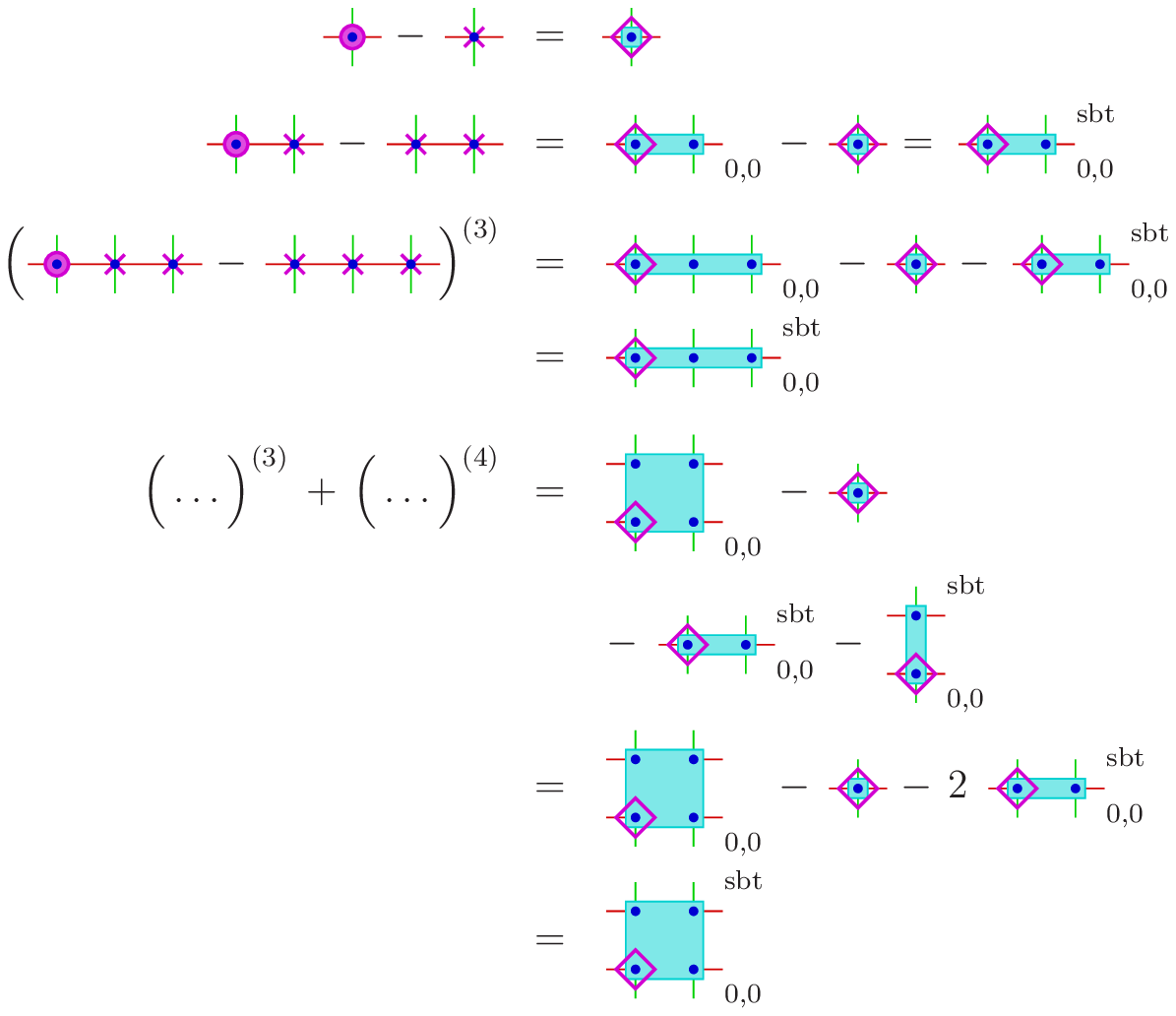}
  \caption{ (Color online)  Inversion of relations shown in
  Fig.~\ref{fig:hoppings_matrix_elements_00}. The exponent sbt (for subtracted) refers to the subcluster subtraction.}
  \label{fig:hoppings_subtraction_00}
\end{figure}

\begin{figure}[t]
  \includegraphics[width=\columnwidth ]{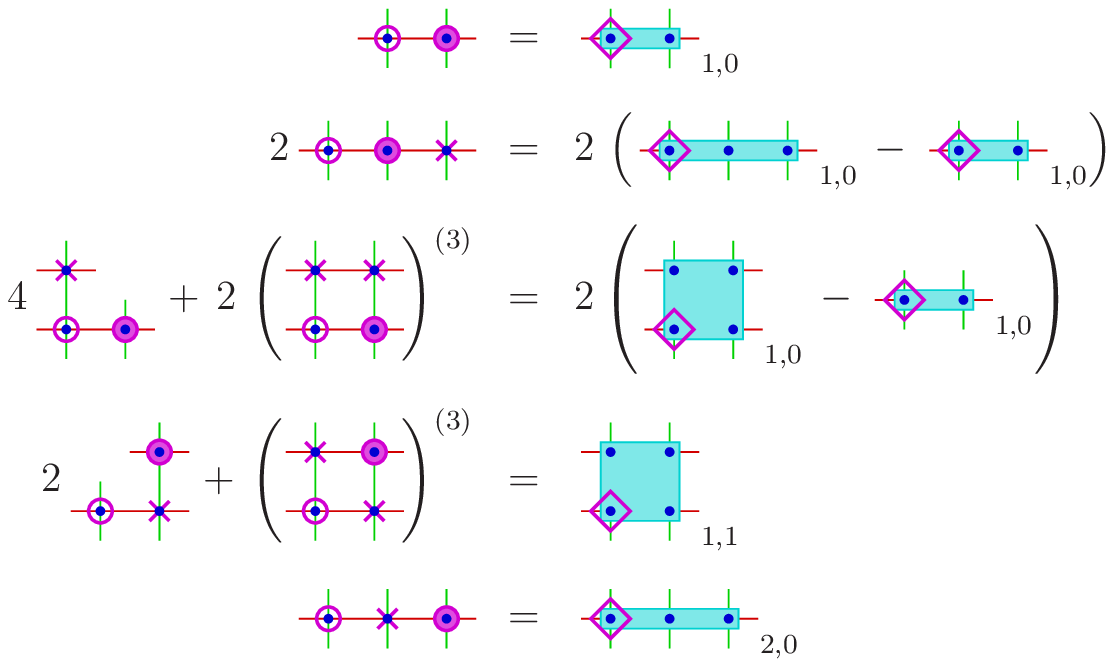}
  \caption{ (Color online)  Inversion of relations shown in 
    Fig.~\ref{fig:hoppings_matrix_elements_n00}.}
  \label{fig:hoppings_subtraction_n00}
\end{figure}

Various contributions of Fig.~\ref{fig:hoppings} can be extracted from
the effective Hamiltonian's matrix elements which are shown in
Figs.~\ref{fig:hoppings_matrix_elements_00} and
\ref{fig:hoppings_matrix_elements_n00}. Again, one can invert these relations,
which leads to a recursive subcluster subtraction scheme (see
Figs.~\ref{fig:hoppings_subtraction_00} and \ref{fig:hoppings_subtraction_n00}).
It is not as easy as in the 0QP sector to write down a concise formula similar
to Eqs.(\ref{eq:FL_sbt}) and (\ref{eq:FL_e0}), because the particle position has
to be taken in consideration, and implies some more restrictions. However things
should be clear from Figs.~\ref{fig:hoppings_matrix_elements_00} and
\ref{fig:hoppings_matrix_elements_n00}. In practice, it is much easier to use a
computer program that determines all relevant subclusters and subtracts their
contributions.

%
%

%
%

%
%

As a final word, let us mention that the 2QP sector can be treated in the same
way since the bound states in the TFIM or the Heisenberg model, that are of
interest to us, behave as a single particle (they are made of two
nearest-neighbor magnons). If one was interested in the 2QP sector where
the two magnons can move freely, one would not only need to perform the 0QP
subtraction as previously, but one would also have to subtract the two 1QP
contributions, in order to extract the 2QP interactions.
We shall now apply this formalism to both systems introduced in Sec.~\ref{sec:models_methods}.

\section{Results for TFIM}
\label{sec:results_TFIM}

\subsection{Series expansion of the low-energy spectrum}

As we have seen, PCUTs provide, order by order, an effective Hamiltonian
$\widetilde{H}^\mathrm{TFIM}_\mathrm{eff}$ which is unitarily equivalent to
$\widetilde{H}_\mathrm{TFIM}$ and commutes with the operator $Q$ counting the
number of antiferromagnetic bonds. One must then analyze
$\widetilde{H}^\mathrm{TFIM}_\mathrm{eff}$ in each sector with a given number
$q$ of antiferromagnetic bonds. In the present work, we derived this effective
Hamiltonian in the sector $q=0$ (0qp or 0QP) up to order 14 while we reached
order 12 for $q=4$ (4qp corresponding to 1QP) and $q=6$ (6qp corresponding to a
bound state of 2QP).
We restrict ourselves to the ``physical" subspace (\textit{i.e.}, the states
corresponding to magnons) where all conserved quantities ($B_p$'s) of
Sec.~\ref{sec:sub:symm_count} are set to $+1$, and postpone the discussion of
``unphysical" states (having antiferromagnetic bonds not corresponding to
a magnon configuration of the initial models) to Sec.~\ref{sec:TC}.

For $q=0$, we obtain the ground-state energy per bond
%
%
\begin{eqnarray}
\label{eq:gse_TFIM}
 e_0 &=& -\frac{1}{2}-\frac{1}{8}\,h^{2}-\frac{1}{384}\,h^{4}-\frac{1}{6144}\,h^{6} -\frac{181}{3538944}\,h^{8}\nonumber\\
     &&-\frac{1388129}{254803968000}\,h^{10} -\frac{67647506447}{25684239974400000}\,h^{12}\nonumber\\
     &&-\frac{707258321166713}{2588971389419520000000}\,h^{14},
\end{eqnarray} 
%
%
which matches with the result given in Ref.~{\cite{Oitmaa91}}. 

The first nontrivial ``physical" sector for the TFIM is the one-magnon sector,
which is a peculiar case of $q=4$, since the four antiferromagnetic bonds must
have a relative position similar to the one shown in
Fig.~\ref{fig:ferro_2flip_far}. We computed the dispersion
$\varepsilon_4(k_x,k_y)$ up to order 12. The list of all relevant hopping
amplitudes is given in Appendix \ref{app:sub:TFIM_hopping1}. Noting that
$\varepsilon_4(k_x,k_y)$ is minimal for $(k_x,k_y)=(0,0)$, one gets the
one-magnon gap
%
\begin{eqnarray}
\Delta_4 &=& 4-\frac{3}{2}\,h^{2}+\frac{43}{96}\,h^{4}-\frac{19993}{27648}\,h^{6}\nonumber\\
         &&+\frac{82873487}{79626240}\,h^{8}-\frac{1901437203257}{1146617856000}\,h^{10}\nonumber\\
         &&+\frac{64764934458802909}{23115815976960000}\,h^{12}.
\end{eqnarray} 
%
%
Note that our results also match those given in Refs.~\onlinecite{Oitmaa06} and \onlinecite{Oitmaa91}
up to some trivial rescaling of the Hamiltonian's parameters.

The next ``physical" sector for the TFIM is a subspace of the 6qp sector which 
corresponds to the bound states discussed above and is illustrated in
Fig.~\ref{fig:ferro_2flip_close}. In this sector, the effective Hamiltonian
describes the hopping of the bound state, whose center of mass lives on the
square lattice formed by the middles of the bonds (dots in
Fig.~\ref{fig:lattice2}). \textit{However, one must state that there are two different types of sites in this lattice} (see Fig.~\ref{fig:lattice2}) 
since bonds can be vertical or horizontal. This is obvious in
Fig.~\ref{fig:ferro_2flip_close} where the two bound states involve a different
pattern of antiferromagnetic bonds. Series expansion of the hopping amplitudes
are given in Appendix \ref{app:sub:TFIM_hopping2} up to
order 12.  From these amplitudes it is clear that the bound state does not have
the same probability to hop in a given direction, for instance $x$, if it is on
a horizontal or on a vertical bond. Of course, this important distinction also
holds for the XXZ model discussed in the next section.

Therefore, one is led to diagonalize a $2\times 2$ matrix for each value of the
center of mass momentum $\boldsymbol{k}=(k_x,k_y)$. One can in particular
extract the gap, which is found to be the lowest of the two energies at
$\boldsymbol{k}=(0,0)$, and reads
%
\begin{eqnarray}
  \Delta_6^- &=& 6-\frac{275}{96}\,h^{4}-\frac{11521}{27648}\,h^{6}
  +\frac{16400551}{7962624}\,h^{8}\\
  && +\frac{1459322986427}{143327232000}\,h^{10}
  -\frac{101780777359633847}{28894769971200000}\,h^{12}.\nonumber
  \end{eqnarray} 
%

The energy of the second mode at zero center of mass momentum is higher and given by
%
\begin{eqnarray}
  \Delta_6^+ &=& 6-\frac{11}{96}\,h^{4}-\frac{115}{1024}\,h^{6}
  -\frac{4956689}{39813120}\,h^{8}\\
  && -\frac{1720028423}{17915904000}\,h^{10}
  -\frac{880952915946869}{9631589990400000}\,h^{12}.\nonumber 
\end{eqnarray} 
%
We do not analyze the associated bound state in detail, but just
note that this mode decays into the two-magnon continuum well before the
critical point. Physical implications of such a decay will be discussed for the
XXZ model below. 

\subsection{Exact diagonalization} \label{sec:sub:ED} 

%
%

To have some finite-size crosschecks of the validity of the
perturbative expansions and the mapping to the Kitaev-type
Hamiltonians, performing exact-diagonalization (ED) studies on the
microscopic models defined above demands particular
adjustment. First, the appearing interactions are of four-body type for
TFIM in Eq.~(\ref{eq:hamTFIM_mapped}) and of seven-body type for XXZ in
Eq.~(\ref{eq:hamXXZ_mapped}). Second, the amount of symmetries is
enormous, which enables us to go to big systems as up to $N=50$
sites. This, however, demands a suitable way to generate the
symmetrically reduced basis, as a loop through all possible spin
configurations would be impracticable.  To overcome the first problem,
we apply the Kernel sweeping method to efficiently implement many-body
interaction terms, where details are elaborated on
in Ref.~\onlinecite{Thomale09_1}. Basically, a small $m$-site Hamiltonian of
the $m$-body interaction is defined, and implemented to sweep over the
whole lattice basis. To address the second point, we do not loop over
all possible $N$-site configurations and subsequently sort out by
symmetry constraints, but start with an allowed state of the basis and
iteratively act on it by the respective Hamilton operator. In each
step, new unprecedented scattering states are collected and added to
the yet incomplete basis, until a new action of $H$ does not produce
new scattering states anymore. In doing so, as $H$ commutes with the
${\mathbb Z}_2$ symmetries Eqs.~(\ref{eq:Bp}) and (\ref{eq:C1C2}), we make sure that
the iteration procedure only acts within the ${\mathbb Z}_2$ charge subsector we
want to consider. For $N=50$, this yields a subblock dimension of
$2^{50/2-1}=16777216$ as alluded to previously, which is already in
suitable range for Lanczos diagonalization algorithms. Still, from
there, we can further exploit lattice symmetries, such as the
translational invariance of the models, to specify the point of the
Brillouin zone we want to study. For the Hamiltonians considered, we
compute the low-energy spectra for clusters up to $N=50$ sites. As
shown in Fig. 15, the finite-size corrections do not allow us to
adequately describe the vicinity of the intermediate and large $h$ limit,
as the proximity of the continuum and the bound-state modes becomes
comparable to the finite-size splitting scale. Apparently, Fig. 15
explicates that the ED data for the largest available system size
corresponds to the expansion data up to $h \sim 0.6$. However, in the
regime where the perturbative expansion data likewise starts to
fluctuate a lot depending on the order of the expansion, the ED data cannot be used for suitable analysis. In the
following, it is thus implicitly assumed that we use exact
diagonalization to crosscheck the implementation of the perturbative
expansions, but can only rely on the latter to highest order to study
the interesting regimes where the bound-state modes approach the
continuum.

\subsection{Analysis and gap ratio at the critical point}

As already discussed in Ref.~\onlinecite{Oitmaa91}, the series expansion for
$\Delta_4$ is strongly diverging so that one has to perform some resummation to
extract gap values for finite $h$. This also seems to be the case for
$\Delta_6^-$, although we only have coefficients up to order 12 for this
quantity. Unfortunately, standard Pad\'{e}-type resummation procedure using the
order 20 expansion for $\Delta_4$ given in Ref.~\onlinecite{Oitmaa91} leads to a
critical point which is still far from the most accurate value. Indeed, naive
Pad\'{e} approximants $([10,10], [8,12], [12,8])$ lead to \mbox{$h_c=(1.65045, 1.57029,
1.73306)$}, respectively, whereas the position of the critical point computed from
series expansion in the high-field phase \cite{He90} is $h_c=1.5216(6)$. Note
that for its classical counterpart which is the critical temperature, higher
precision of order $10^{-7}$ has been reached (see, for instance, Ref.~\onlinecite{Arisue04}
and references therein).

Here, our aim is to check a result coming from field-theoretical calculations
performed by Caselle {\it et al.} \cite{Caselle99,Caselle02} who predict that
$\Delta_6^-/\Delta_4|_{h=h_c}\simeq 1.8$. Recently, this prediction has been
improved using numerical diagonalization \cite{Nishiyama08}, giving a ratio of
$1.84(3)$.

Given that the transition point is defined by \mbox{$\Delta_4=0$}, a finite ratio at
the critical field means that $\textit{(i)}$ $\Delta_6^-=0$ at
this point, and $\textit{(ii)}$ $\Delta_6^-$ vanishes with the same critical exponent
$\nu$. It can therefore be expected that the direct extrapolation of
$\Delta_6^-$ is at least as complicated as the one for $\Delta_4$ which is actually the case.
By contrast, one may hope that the ratio $\Delta_6^-/\Delta_4$, on which we focus below, has a better behavior.

\begin{figure}[t]
  \includegraphics[width=\columnwidth ]{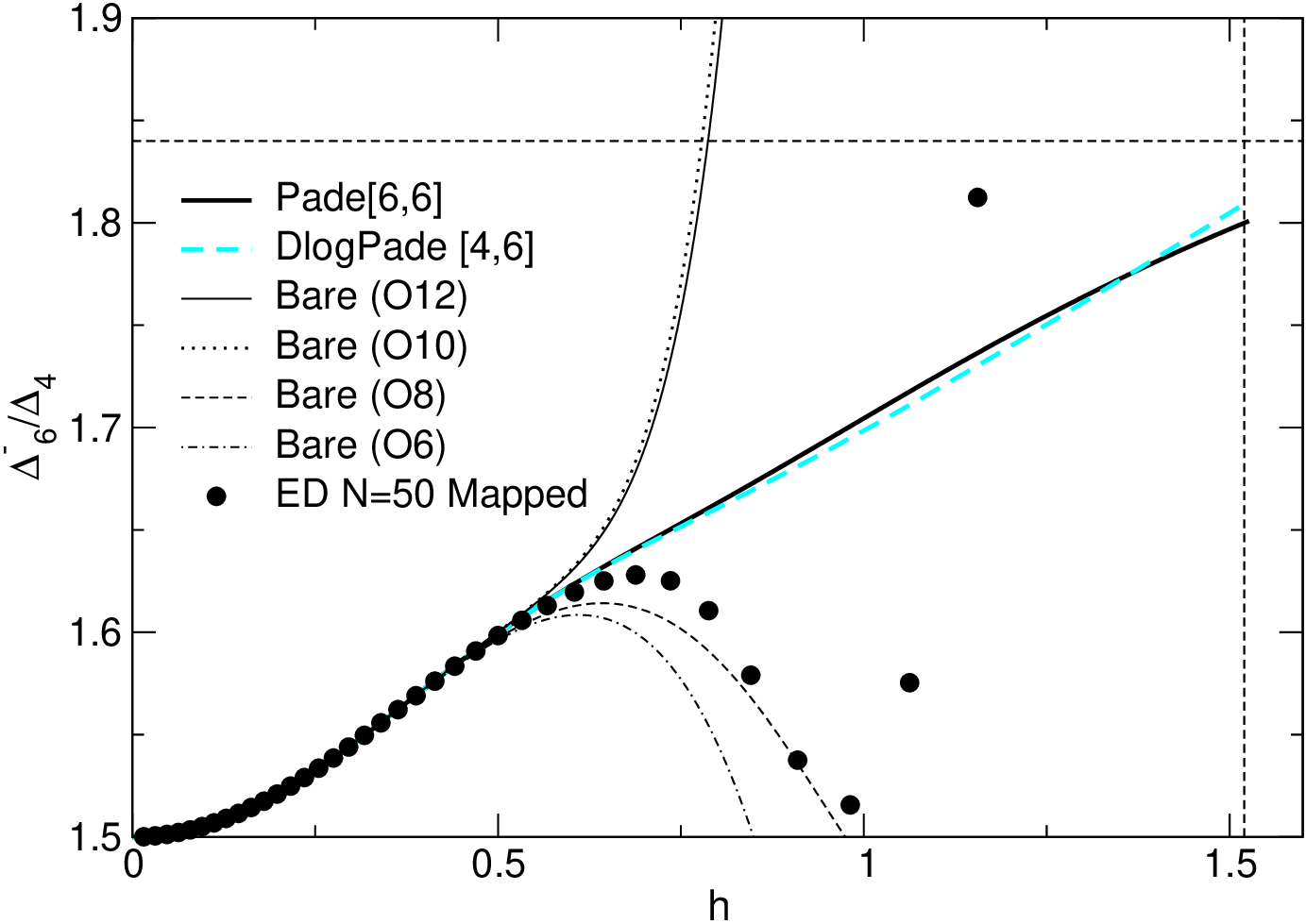}
  \caption{ (Color online)  Thin lines represent the bare ratio
  $\Delta_6^-/\Delta_4$ for different maximal orders 6, 8, 10, and 12 as a function
  of the magnetic field $h$. Thick lines correspond to different approximations
  of this ratio. Filled circles denote ED data of the mapped
  model (\ref{eq:hamTFIM_mapped}) with $N=50$ sites (bonds of the original lattice). Dashed vertical
  (horizontal) line marks the value of the magnetic field (ratio) at the
  critical point as obtained from the numerical diagonalizations of
  Ref.~\onlinecite{Nishiyama08}.}
  \label{fig:ratio_ising}
\end{figure}

The bare series of the ratio $\Delta_6^-/\Delta_4$ up to order 12 reads
%
\begin{eqnarray}
  \frac{\Delta_6^-}{\Delta_4} &=& \frac{3}{2}+\frac{9}{16}\,h^{2}
  -\frac{517}{768}\,h^{4}-\frac{32831}{221184}\,h^{6} \\
  && +\frac{156729359}{637009920}\,h^{8}
  +\frac{27593405457803}{9172942848000}\,h^{10}\nonumber \\
  && -\frac{415396528829457211}{924632639078400000}\,h^{12},\nonumber
\end{eqnarray} 
%
and is shown in Fig.~\ref{fig:ratio_ising}. Clearly, the bare series is still
alternating and the convergence is rather poor ($h\le 0.5$) so extrapolation
schemes are mandatory. 
%
%
First we tried standard Pad\'{e} approximants. We found that all approximants
$[n,m]$ with $n+m\le10$ give no useful result in the sense that the approximant
either has a spurious pole or shows a diverging behavior well before the
critical field $h_{c}$. Looking at the Pad\'{e} approximants with $n+m=12$ we
found only two valid cases, namely, $[8,4]$ and $[6,6]$. Still no converging
picture emerges because the $[8,4]$ approximant displays a diverging behavior.
However,  the approximant $[6,6]$ is the first one to behave smoothly (see
Fig.~\ref{fig:ratio_ising}). We conclude that the Pad\'{e} analysis gives no
convergent picture and it seems that one needs at least the order 12 series to
catch the physics of this ratio close to the critical point.

To proceed further, we used DlogPad\'{e} extrapolation which is usually more reliable than
Pad\'{e} extrapolations for positive quantities. Among all approximants with $n\ge4$ and $m\ge4$, 
DlogPad\'{e} $[4,6]$ is the only one that has no spurious pole.  This extrapolation is shown in
Fig.~\ref{fig:ratio_ising}. It can be seen that the ratio seems to behave almost
linearly as a function of the field close to the critical point. The ratio at
the critical field $h_c$ is $1.81$ which is very close to the
numerical value \cite{Nishiyama08}. So one finds again that one needs at least order 12
to capture the expected behavior. Clearly, higher orders are expected to give
more valuable insights in this quantity but these are beyond the scope of this
work. 

The relevance of higher orders can be also understood from the fact that
fluctuations on rather large length scales are required to follow this ratio up
to the critical point. This is in agreement with ED results, displayed in Fig.~\ref{fig:ratio_ising},
which are qualitatively similar to the bare PCUTs series.

\section{Results for the XXZ model}
\label{sec:results_XXZ}

\subsection{Series expansion of the low-energy spectrum}

Similarly, we computed the low-energy spectrum of the XXZ model by
the PCUTs method. As previously, we restrict the discussion again to the ``physical"
subspaces $q\in\{0,4,6\}$ corresponding to the ground state, to \mbox{one-magnon}
states and to two-magnon bound states.
We derive the effective Hamiltonian up to order $10$ in the 0QP sector and up
to order $8$ in the 1QP and 2QP sectors. 

For $q=0$, we obtain the ground-state energy per dimer
%
\begin{eqnarray}
  e_0 &=& -\frac{1}{2}-\frac{1}{6}\,\lambda^{2}+\frac{1}{1080}\,\lambda^{4}
  -\frac{3587}{2268000}\,\lambda^{6}\\
  && -\frac{660294389}{800150400000}\,\lambda^{8}
  -\frac{156875294970593831}{503046234915840000000}\,\lambda^{10},\nonumber
\end{eqnarray} 
%
which matches with the result given in Ref.~\onlinecite{Singh89}.

The one-magnon sector corresponds again to the first nontrivial ``physical"
sector with $q=4$. The four antiferromagnetic bonds must remain in a
closed-pack relative position such that they share a site in the original square lattice (see Fig.~\ref{fig:ferro_2flip_far}). The dispersion can be obtained from the list of hopping amplitudes given
in Appendix \ref{app:sub:XXZ_hopping1}. The minimum of the dispersion is found
at $\boldsymbol{k}=(0,0)$, and one gets for the one-magnon gap, in accordance
with Ref.~\onlinecite{Zheng91},
%
\begin{eqnarray}
  \Delta_4 &=& 4-\frac{10}{3}\,\lambda^{2}+\frac{137}{216}\,\lambda^{4}
  -\frac{13039847}{15552000}\,\lambda^{6}\nonumber\\
  && +\frac{124898889761701}{230443315200000}\,\lambda^{8}.
\end{eqnarray} 
%
%

We now switch to the $q=6$ sector that corresponds to the two-magnon
bound state. Its analysis follows exactly the same steps as  for the TFIM discussed in the previous section. 
Series expansions of the hopping amplitudes for the bound state are given in
Appendix \ref{app:sub:XXZ_hopping2} up to order 8. One has to diagonalize a $2\times 2$ matrix for each value of the
center-of-mass momentum $\boldsymbol{k}=(k_x,k_y)$. The gap is found at
$\boldsymbol{k}=(0,0)$, and reads
%
\begin{eqnarray}
  \Delta_6^- &=& 6-\frac{10}{3}\,\lambda^{2}+\frac{323}{540}\,\lambda^{4}
  -\frac{1435321}{324000}\,\lambda^{6}\nonumber\\
  && +\frac{3809941658983}{320060160000}\,\lambda^{8}.
\end{eqnarray} 
%

The other bound mode at higher energy with the same momentum $\boldsymbol{k}=(0,0)$
has the expansion
%
\begin{eqnarray}
  \Delta_6^+ &=& 6+\frac{2}{3}\,\lambda^{2}-\frac{619}{1080}\,\lambda^{4}
  -\frac{482989}{1036800}\,\lambda^{6}\nonumber\\
  && -\frac{15320370383651}{19203609600000}\,\lambda^{8}.
\end{eqnarray} 
%
We would like to point out that the two different orientations of the bound
state are missed in Ref.~\onlinecite{Hamer09}. As a consequence, the dispersion
for arbitrary momentum of the bound state is not correct in this reference. Only the gap in this
sector matches with our results (which certainly means that all hopping
amplitudes in Ref.~\onlinecite{Hamer09} are correct).

\subsection{Fate of the bound states}

An interesting question is to determine when the two-magnon bound states
decay as a function of $\lambda$. To this end, we first focus on the case of
total momentum $\boldsymbol{k}=(0,0)$, describing  low-energy physics. Bound states decay for this momentum at latest at
the Heisenberg point $\lambda=1$, where the one-magnon gap closes and therefore
all multimagnon continua have a gapless spectrum. 

The lowest energy of the two-magnon continuum is found at total momentum $\boldsymbol{k}=(0,0)$ and is given by twice the one-magnon gap $\Delta_4$. Bound states start to acquire a finite
lifetime once their energy is degenerate with the lower band edge of the
two-magnon continuum. Consequently, we determine the values of $\lambda$ for
which ratios $2\Delta_4/\Delta_6^\pm$ are equal to one.
In the following, we restrict the discussion to the PCUTs results since reliable ED results would require enormous system sizes (with 36 sites one would only capture terms of order $\lambda^2$). 
The bound state is indeed an extended object and the $\lambda$ term in the XXZ Hamiltonian (\ref{eq:hamXXZ_3}) involves nearest-neighbor interactions, contrary to the purely local $h$ term in the TFIM Hamiltonian (\ref{eq:hamTFIM}).

%
\begin{figure}[t]
  \includegraphics[width=\columnwidth ]{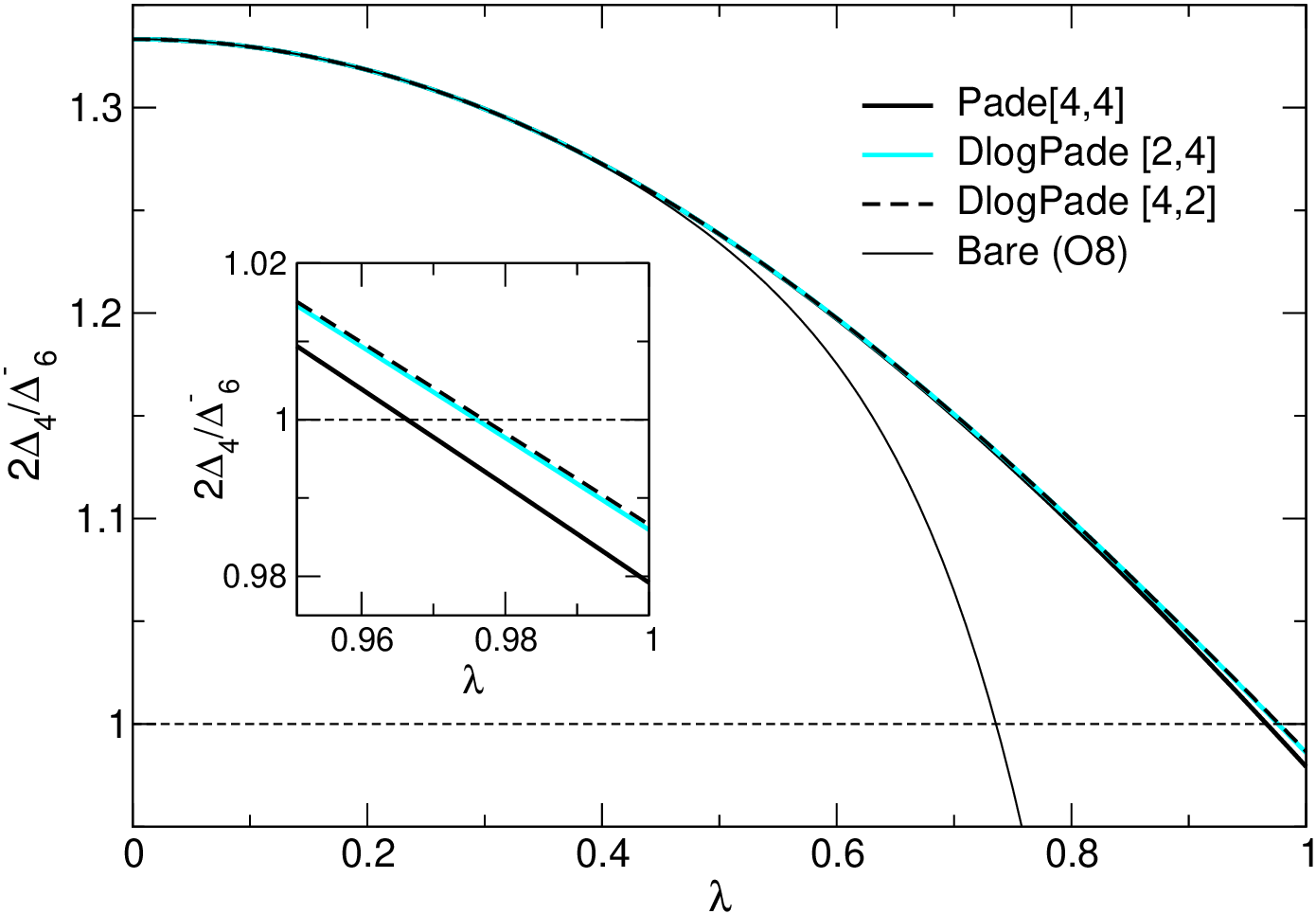}
  \caption{ (Color online)  Thin blue line represents the bare ratio
  $2\Delta_4/\Delta_6^-$ for the maximal order 8 as a function of the anisotropy parameter
  $\lambda$. Thick lines correspond to different approximations of this ratio.
  The dashed horizontal line marks the value 1 of the ratio. The value
  $\lambda=1$ corresponds to the Heisenberg point.}
  \label{fig:ratio_xxz}
\end{figure}
%

Obviously, the high-energy mode $\Delta_6^+$ decays first. Using
different extrapolation schemes such as Pad\'{e} and DlogPad\'{e}, we find 
that it disappears  for $\lambda\simeq 0.5401(1)$, \textit{i.~e.}, for a rather small value
which explains the high accuracy. Beyond this point, the bound
state gains a finite lifetime and, strictly speaking, a perturbative
derivation of a block-diagonal Hamiltonian becomes
impossible \cite{Fischer09}. One can expect that the decayed bound state shows up as resonances
inside the continua of dynamical correlation functions. The two-magnon peak
observed in the theoretical Raman response at the Heisenberg point $\lambda=1$
(Refs.~\onlinecite{Singh89_2} and \onlinecite{Canali92}) and experimentally detected in the undoped cuprate
compounds \cite{Lyons89,Ohana89} might be a remnant of this bound state. However,  a
correct physical description of the decay process is beyond any series expansion
study. This scenario is further confirmed by the fact that any Pad\'{e}
extrapolation of the energy $\Delta_6^+$ has poles in the denominator.

Concerning the low-energy properties, in analogy to the Ising case studied in
the previous section, the fate of the low-energy mode $\Delta_6^-$ is more
interesting but also more challenging. The bare ratio $2\Delta_4/\Delta_6^-$
together with different approximants are shown in Fig.~\ref{fig:ratio_xxz}. We observe
that no pole shows up in these approximants. 
Consistently, all approximations indicate that the decay takes place very close to the Heisenberg
point ($\lambda \simeq 0.97$). 
Taking into account that the series have been obtained up to order 8 only and have only even orders, one cannot tell precisely whether the merging point is located exactly at the critical point ($\lambda=1$), just as
in the TFIM, or not. One argument in favor of the former scenario is that,
for any finite-size system, the ground state of
the SU(2)-invariant Heisenberg point is a singlet and the gapped elementary excitation
is a threefold degenerate triplet, with total magnetization $S_z\in\{-1,0,+1\}$.
The excitations with $S_z\in\{-1,+1\}$ can be identified with one-magnon excitations found
for $\lambda<1$, whereas the excitation with $S_z=0$ has to be a two-magnon bound state. 
The gap of these excitations should furthermore vanish in the thermodynamical limit considered in the present PCUTs study.

\section{The toric code model in a transverse field with $\mathbf{J_ \mathrm{p}=0}$}
\label{sec:TC}

As mentioned in Sec.~\ref{sec:sub:symm_count}, the toric code Hamiltonian (\ref{eq:hamTFIM_mapped}) deserves to be analyzed on its own. Indeed, in the original toric code model \cite{Kitaev03}, Kitaev focused on  the isotropic coupling $J_\mathrm{s}=J_\mathrm{p}$ [see Eq.~(\ref{eq:hamTC}) for notations]. Here, the bond description of the TFIM leads us to consider a different situation where ({\it i}) $J_ \mathrm{p}=0$ and ({\it ii})  a magnetic field in the $z$ direction is introduced. Let us stress that such a model is very close to the Xu-Moore model \cite{Xu04} and, in some sense, very similar to the parallel-field problem discussed in 
Refs.~\onlinecite{Trebst07} and \onlinecite{Vidal09_1,Hamma08,Tupitsyn08}. Most importantly, it is exactly the model introduced by Wegner in his seminal paper \cite{Wegner71} (see also Ref.~\onlinecite{Hamma08}) but, here, $B_p$ operators are conserved quantities that can be in any configuration. This crucial difference raises several questions that we shall address in the present section.

\begin{figure}[t]
  \includegraphics[width=0.5\columnwidth ]{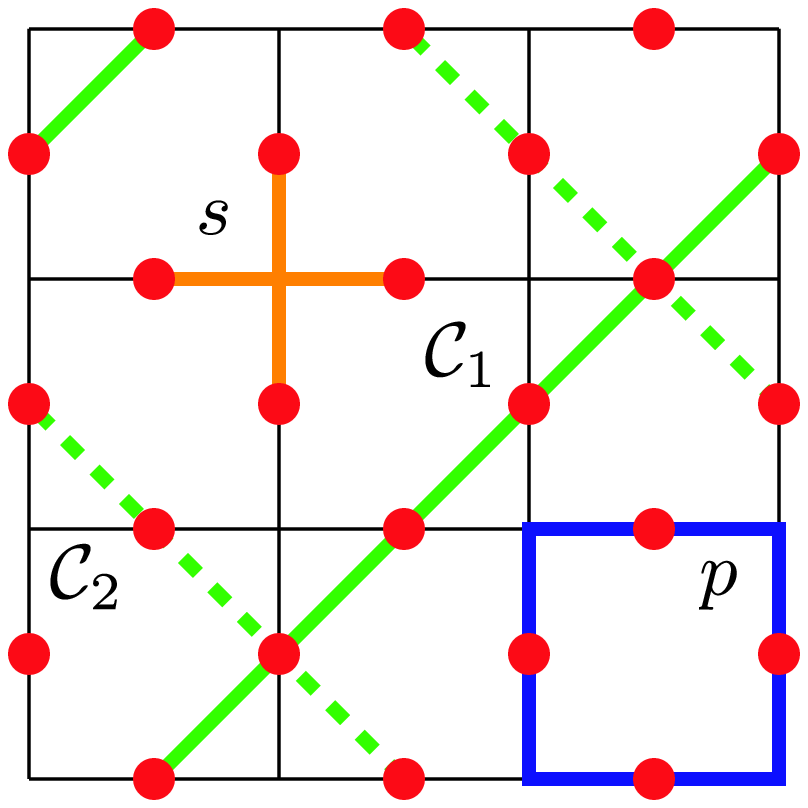}
  \caption{ (Color online)  Square lattice on which the toric code Hamiltonian (\ref{eq:hamTCF}) is defined with periodic boundary conditions. Sites are represented with dots. We also show a vertex $s$, a plaquette $p$, a diagonal contour $\mathcal{C}_1$, and an antidiagonal contour $\mathcal{C}_2$, used to define various operators (see text).}
  \label{fig:lattice3}
\end{figure}

Let us first rewrite the Hamiltonian and various operators of Sec.~\ref{sec:sub:symm_count} in the toric code language. The Hamiltonian reads

%
\begin{equation}
\label{eq:hamTCF}
  H = - J \sum_s A_s - h \sum_i \sigma_i^z,
  \end{equation}
%
where the spins live on the bonds of a square lattice (see Fig.~\ref{fig:lattice3}) and the $A_s=\prod_{i\in s}\sigma^x_i$ operators involve the four sites around a vertex $s$. The plaquette operators $B_p=\prod_{i\in p}\sigma^z_i$ are conserved. For a system defined with periodic boundary conditions the cycle operators $\prod_{i\in\mathcal{C}}\sigma^z_i$ defined on diagonal or antidiagonal contours, such as those shown in Fig.~\ref{fig:lattice3}, are conserved as well.

%
%

The main difference with the Xu-Moore Hamiltonian \cite{Xu04} is that $A_s$ operators only act on vertices of the square lattice and not on plaquettes.
Furthermore, in the Xu-Moore model, the cycle operators are still conserved, contrary to the $B_p$ operators.
The dynamics of the quasiparticles is thus expected to be more constrained than in the Xu-Moore model which already exhibits dimensional reduction.

In the following, we shall discuss separately the two limits (large and small $J/h$) for which we computed perturbatively the low-energy spectrum. One can already note that in the small-field limit, the system is in a topological phase ({\it i.~e.}, the ground-state degeneracy depends on the surface genus) whereas at large field the ground state is obviously unique. Thus, one expects a (topological) quantum phase transition when varying the ratio $J/h$. 

%
\subsection{Large-field limit $h\gg J$}
\label{sec:TCLF}
%

For $J=0$, elementary excitations are usual magnons obtained by flipping any spins from the ground state which is the state fully polarized in the field ($z$) direction. PCUTs formalism will then give a dressed magnon description when switching on $J$ \cite{Vidal09_1}. 
Setting $h=1/2$, the ground-state energy per bond (which coincides with the 0qp level) is obtained from 
Eq.~(\ref{eq:gse_TFIM}) by replacing $h$ by $J$. 

However, for this model, arbitrary qp sectors are allowed although conservation of $B_p$'s imposes severe constraints on the spectrum. For $q=1$, one only has a nondispersive (excited) level at energy
%
\begin{eqnarray}
  \Delta_1 &=& 1-\frac{J^2}{2}+\frac{3 J^4}{32}-\frac{31 J^6}{768}
  +\frac{299233 J^8}{15925248} \nonumber \\
  && -\frac{2014178639 J^{10}}{764411904000},
\end{eqnarray} 
%
since any displacement of this localized dressed magnon would modify the $B_p$'s configuration. 

\begin{figure}[t]
  \includegraphics[width=0.5\columnwidth ]{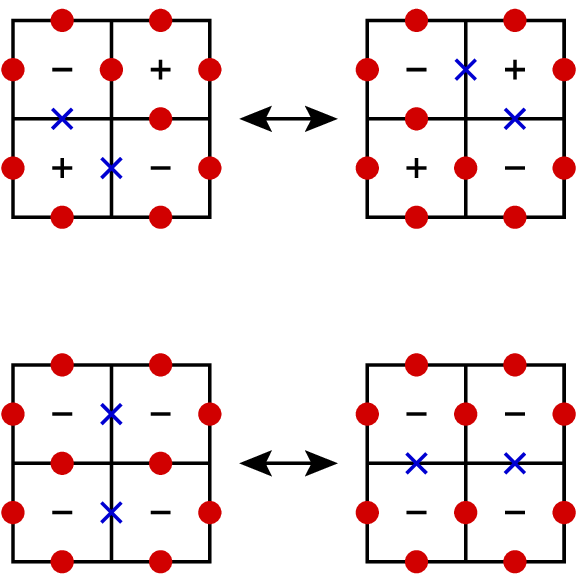}
  \caption{ (Color online)  Illustration of the two-magnon dynamics in the high-field limit of Hamiltonian
(\ref{eq:hamTCF}). Dressed magnons are depicted as crosses. The $+$ and $-$ signs refer to the eigenvalue $+1$ or $-1$ of the plaquette operators $B_p$.}
  \label{fig:dynamics_2qp}
\end{figure}

For $q=2$, the only possible dynamics is provided by some flip-flop
processes depicted in Fig.~\ref{fig:dynamics_2qp} (we do not consider the case where the two dressed magnons are far apart).
Both configurations shown in Fig.~\ref{fig:dynamics_2qp} yield the same gap, which reads
%
\begin{eqnarray}
  \Delta_2 &
  =& 2-J-\frac{J^2}{4}+\frac{J^3}{8}-\frac{13 J^4}{192}+\frac{29 J^5}{768}
  -\frac{445 J^6}{13824} \nonumber \\
  && +\frac{739J^7}{36864}-\frac{608839 J^8}{79626240}
  -\frac{2462069 J^9}{9555148800} \nonumber \\
  && +\frac{21097903 J^{10}}{152882380800}.
\end{eqnarray} 
%
Note that in the Xu-Moore model, the pair of nearest-neighbor magnons (top of Fig.~\ref{fig:dynamics_2qp}) is allowed to hop at any distance (but in a one-dimensional stripe) \cite{Vidal09_2}.

%
%

We have not computed energies of states with more magnons. Let us simply mention the following two points.
For $q=3$, the problem is similar to $q=2$ and no dynamics of the QPs is allowed apart from flip-flip processes. 
The first dispersive mode is the one involving four magnons, just as in the Xu-Moore model \cite{Vidal09_2}.

%
\subsection{Small-field limit $h\ll J$}
\label{sec:TCSF}
%

Let us now turn to the opposite limit where $J$ is much larger than $h$ and for which the system is in a topological phase. For convenience, let us set $J=1/2$. Since the $B_p$ operators do not appear in the Hamiltonian, all states are macroscopically degenerate when $h=0$. It is thus natural to wonder what the effect of the perturbation is. A calculation at fourth order of the effective Hamiltonian in the sector where there is no effective vertex excitation (still using the PCUTs formalism) yields

%
\begin{equation}
	H_\mathrm{eff} = -N\left(\frac{1}{4}+\frac{h_z^2}{2}+\frac{5h_z^4}{8}\right)-\frac{5h_z^4}{2}\sum_p B_p,
\end{equation}
%
where $N$ is the number of sites (and thus twice the number of plaquettes or of vertices). We thus see that, once the magnetic field is switched on, the ground state belongs to the sector where each operator $B_p$ has eigenvalue $+1$. As a check, note that setting all $B_p$ operators to one in the above formula gives the ground-state energy, which matches with Eq.~(8) of Ref.~\onlinecite{Vidal09_1} (apart from a trivial constant shift since no $B_p$ operator appears in our Hamiltonian, and after setting $h_x$ to zero in this equation).
The fact that the sector with no plaquette excitation is selected by the perturbation is consistent with an Ising-type  quantum phase transition. Indeed, in this sector, the model with Hamiltonian (\ref{eq:hamTCF}) is dual to the two-dimensional transverse-field Ising model (which follows from the same arguments as those used in Ref.~\onlinecite{Trebst07}).

\section{Conclusions}
\label{sec:C}

We have studied two-magnon bound states in the TFIM and XXZ models  using PCUTs  around the Ising limit. This method gives a nice physical picture of these excitations in terms of dressed magnons which can be of two kinds on a square lattice depending on the "frustrated bond" localizing it (horizontal or vertical). Consequently, one obtains two different bound modes contrary to previous claims \cite{Hamer09,Oguchi73}. 
For both systems, we find that the lowest-energy gap vanishes at the critical point, or at least in a neighborhood that cannot be distinguished from this point with actual series expansions. Solving this issue, especially for the XXZ model,  would clearly require much higher orders or a nonperturbative treatment as the one considered in Refs.~\onlinecite{Caselle99} and \onlinecite{Caselle02} for the TFIM. 

From a methodological point of view, we have adapted Entings' finite-lattice method commonly used in statistical mechanics to quantum problems. This approach basically consists in a cluster embedding which considerably increases the efficiency (from the time and memory point of view) of the PCUTs method we used here. Note that for the XXZ model, such an improvement allowed us to reach the same maximum order as standard series expansions techniques (see Refs.~\onlinecite{Oitmaa06} and \onlinecite{Hamer09}) based on a more sophisticated graph analysis. Clearly, adapting such a graph description to PCUTs is a crucial issue which is currently under study and should allow one to reach higher orders.


\acknowledgments We wish to thank M. Caselle, M.~D. Schulz, and G.~S. Uhrig for fruitful discussions. K.P.S. acknowledges ESF and EuroHorcs for funding through his EURYI. 

\appendix

\section{Low-energy spectrum of the TFIM}
\label{app:low_energy_TFIM}

\subsection{One-magnon hopping amplitudes}
\label{app:sub:TFIM_hopping1}

One-magnon excitations are located on sites of the square lattice. The
corresponding hopping amplitudes $t_{i,j}$ of $i$ sites in $x$ direction and $j$
sites in $y$ direction are given below. Hopping amplitudes that are not given
can be deduced from the symmetries of the lattice and the hermiticity of the
Hamiltonian.

\begin{eqnarray}
  t_{0,0} &=& 4-\frac{1}{2}\,h^{2}+\frac{19}{96}\,h^{4}-\frac{4745}{27648}\,h^{6}+\frac{15167827}{79626240}\,h^{8}\nonumber\\
         &&-\frac{274582941007}{1146617856000}\,h^{10}+\frac{39052830905417587}{115579079884800000}\,h^{12}\nonumber\\
  t_{1,0} &=& -\frac{1}{4}\,h^{2}+\frac{1}{16}\,h^{4}-\frac{785}{9216}\,h^{6}+\frac{49355}{442368}\,h^{8}\nonumber\\
          &&-\frac{1325086777}{8493465600}\,h^{10}+\frac{503970370332103}{2140353331200000}\,h^{12}\nonumber\\
  t_{1,1} &=& -\frac{29}{768}\,h^{6}+\frac{82181}{1327104}\,h^{8}-\frac{94291093}{955514880}\,h^{10}\nonumber\\
          &&+\frac{620486307800173}{3852635996160000}\,h^{12}\nonumber\\
  t_{2,0} &=& -\frac{7}{768}\,h^{6}+\frac{25789}{1327104}\,h^{8}-\frac{15624304847}{382205952000}\,h^{10}\nonumber\\
      &&+\frac{1492786454328649}{19263179980800000}\,h^{12}\nonumber
   \end{eqnarray}
  \begin{eqnarray}
   t_{2,1} &=& -\frac{23}{9216}\,h^{6}+\frac{11441}{1327104}\,h^{8}-\frac{17711235911}{764411904000}\,h^{10}\nonumber\\
          &&+\frac{273663316786417}{5503765708800000}\,h^{12}\nonumber\\
  t_{3,0} &=& -\frac{23}{27648}\,h^{6}+\frac{1529}{995328}\,h^{8}-\frac{5258321581}{1146617856000}\,h^{10}\nonumber\\
         &&+\frac{202725363389893}{16511297126400000}\,h^{12}\nonumber\\
  t_{2,2} &=& \frac{307}{884736}\,h^{8}-\frac{442674073}{127401984000}\,h^{10}\nonumber\\
          &&+\frac{31180724115977}{2568423997440000}\,h^{12}\nonumber\\
  t_{3,1} &=& \frac{307}{1327104}\,h^{8}-\frac{258078503}{127401984000}\,h^{10}\nonumber\\
          &&+\frac{268024184026579}{38526359961600000}\,h^{12}\nonumber\\
  t_{4,0} &=& \frac{307}{5308416}\,h^{8}-\frac{24494311}{84934656000}\,h^{10}\nonumber\\
         &&+\frac{11181306857333}{11007531417600000}\,h^{12}\nonumber\\
  t_{3,2} &=& -\frac{1621657}{12740198400}\,h^{10}+\frac{359038393}{339738624000}\,h^{12}\nonumber\\
  t_{4,1} &=& -\frac{1621657}{25480396800}\,h^{10}+\frac{54196069259}{122305904640000}\,h^{12}\nonumber\\
  t_{5,0} &=& -\frac{1621657}{127401984000}\,h^{10}+\frac{11111462099}{203843174400000}\,h^{12}\nonumber\\
  t_{3,3} &=& \frac{63560779}{1834588569600}\,h^{12}\nonumber\\
  t_{4,2} &=& \frac{63560779}{2446118092800}\,h^{12}\nonumber\\
  t_{5,1} &=& \frac{63560779}{6115295232000}\,h^{12}\nonumber\\
  t_{6,0} &=& \frac{63560779}{36691771392000}\,h^{12}\nonumber
\end{eqnarray}

\subsection{Two-magnon bound state hopping amplitudes}
\label{app:sub:TFIM_hopping2}

The two-magnon bound states are located at the centers of the bonds on the
square lattice and exist in two kinds~: the bound state can indeed live on a
horizontal or a vertical link denoted $\mathrm{h}$ and $\mathrm{v}$ respectively.
The hopping amplitudes of a horizontally oriented bound state
$t^{\mathrm{h},\alpha}_{i,j}$ are listed below. The corresponding hopping
elements for the vertically oriented bound state can be deduced by reversing the
$x$ and $y$ components, \textit{i.~e.},
$t^{\mathrm{v},\alpha}_{i,j}=t^{\mathrm{h},\alpha}_{j,i}$. Again, the symmetries
of the lattice and the hermiticity of the Hamiltonian have been used to restrict
the number of given amplitudes.

\begin{eqnarray}
  t^{\mathrm{h,h}}_{0,0} &=& 6-\frac{5}{8}\,h^{4}-\frac{8573}{27648}\,h^{6}+\frac{28100449}{79626240}\,h^{8}\nonumber\\
                       &&+\frac{34908351643}{23887872000}\,h^{10}+\frac{8582632522763479}{23115815976960000}\,h^{12}\nonumber
 \end{eqnarray}
\begin{eqnarray}
 t^{\mathrm{h,h}}_{0,1} &=& -\frac{1}{3}\,h^{4}-\frac{1}{12}\,h^{6}+\frac{8209}{41472}\,h^{8}+\frac{4323615563}{4777574400}\,h^{10}\nonumber\\
                       &&+\frac{844706924968673}{8255648563200000}\,h^{12}\nonumber\\
  t^{\mathrm{h,v}}_{\frac{1}{2},\frac{1}{2}} &=& -\frac{1}{3}\,h^{4}-\frac{59}{768}\,h^{6}+\frac{1052251}{4976640}\,h^{8}+\frac{266099856239}{286654464000}\,h^{10}\nonumber\\
                       &&+\frac{7513297407630751}{115579079884800000}\,h^{12}\nonumber\\
  t^{\mathrm{h,h}}_{1,0} &=& -\frac{1}{12}\,h^{4}+\frac{7}{192}\,h^{6}+\frac{102731}{2211840}\,h^{8}+\frac{5273001593}{15925248000}\,h^{10}\nonumber\\
                       &&-\frac{818755629011689}{3210529996800000}\,h^{12}\nonumber\\
  t^{\mathrm{h,h}}_{0,2} &=& \frac{1}{864}\,h^{6}+\frac{207833}{3981312}\,h^{8}+\frac{41166508631}{286654464000}\,h^{10}\nonumber\\
                       &&-\frac{687140651194889}{8255648563200000}\,h^{12}\nonumber\\
  t^{\mathrm{h,v}}_{\frac{1}{2},\frac{3}{2}} &=& -\frac{1}{192}\,h^{4}+\frac{7}{384}\,h^{6}+\frac{94081}{2654208}\,h^{8}+\frac{118209917}{707788800}\,h^{10}\nonumber\\
                       &&-\frac{993567226390117}{6421059993600000}\,h^{12}\nonumber\\
  t^{\mathrm{h,h}}_{1,1} &=& -\frac{1}{192}\,h^{4}+\frac{43}{1536}\,h^{6}+\frac{5731}{294912}\,h^{8}+\frac{556424309}{3185049600}\,h^{10}\nonumber\\
                       &&-\frac{189675771174199}{917294284800000}\,h^{12}\nonumber\\
  t^{\mathrm{h,v}}_{\frac{3}{2},\frac{1}{2}} &=& -\frac{1}{192}\,h^{4}+\frac{7}{384}\,h^{6}+\frac{94081}{2654208}\,h^{8}+\frac{118209917}{707788800}\,h^{10}\nonumber\\
                       &&-\frac{993567226390117}{6421059993600000}\,h^{12}\nonumber\\
  t^{\mathrm{h,h}}_{2,0} &=& -\frac{1}{192}\,h^{4}+\frac{13}{1536}\,h^{6}-\frac{199}{18432}\,h^{8}+\frac{20876171}{1061683200}\,h^{10}\nonumber\\
                       &&-\frac{2349250291300307}{23115815976960000}\,h^{12}\nonumber\\
 t^{\mathrm{h,h}}_{0,3} &=& -\frac{271}{1769472}\,h^{8}+\frac{23677}{19660800}\,h^{10}\nonumber\\
                        &&-\frac{471248557368467}{16511297126400000}\,h^{12}\nonumber\\
  t^{\mathrm{h,v}}_{\frac{1}{2},\frac{5}{2}} &=& \frac{1}{1728}\,h^{6}-\frac{22393}{10616832}\,h^{8}+\frac{12078781789}{2293235712000}\,h^{10}\nonumber\\
                       &&-\frac{235040908701859}{5503765708800000}\,h^{12}\nonumber\\
  t^{\mathrm{h,h}}_{1,2} &=& \frac{1}{1728}\,h^{6}-\frac{15085}{5308416}\,h^{8}+\frac{1160612737}{229323571200}\,h^{10}\nonumber\\
                       &&-\frac{625401171948659}{8255648563200000}\,h^{12}\nonumber\\
  t^{\mathrm{h,v}}_{\frac{3}{2},\frac{3}{2}} &=& \frac{1}{864}\,h^{6}-\frac{5939}{1327104}\,h^{8}+\frac{1534754177}{191102976000}\,h^{10}\nonumber\\
                       &&-\frac{470006557776271}{5778953994240000}\,h^{12}\nonumber\\
  t^{\mathrm{h,h}}_{2,1} &=& \frac{1}{864}\,h^{6}-\frac{5939}{1327104}\,h^{8}+\frac{77266223}{7644119040}\,h^{10}\nonumber\\
                      &&-\frac{23741703266443}{412782428160000}\,h^{12}\nonumber\\
  t^{\mathrm{h,v}}_{\frac{5}{2},\frac{1}{2}} &=& \frac{1}{1728}\,h^{6}-\frac{22393}{10616832}\,h^{8}+\frac{12078781789}{2293235712000}\,h^{10}\nonumber\\
                      &&-\frac{235040908701859}{5503765708800000}\,h^{12}\nonumber
\end{eqnarray}
\begin{eqnarray}
  t^{\mathrm{h,h}}_{3,0} &=& \frac{1}{1728}\,h^{6}-\frac{203}{147456}\,h^{8}+\frac{872395631}{286654464000}\,h^{10}\nonumber\\
                       &&-\frac{95952207356309}{11557907988480000}\,h^{12}\nonumber\\
  t^{\mathrm{h,h}}_{0,4} &=& \frac{7579}{331776000}\,h^{10}-\frac{1916092087}{7644119040000}\,h^{12}\nonumber\\
  t^{\mathrm{h,v}}_{\frac{1}{2},\frac{7}{2}} &=& -\frac{271}{3538944}\,h^{8}+\frac{4822213}{10616832000}\,h^{10}\nonumber\\
                       &&-\frac{143095110961}{73383542784000}\,h^{12}\nonumber\\
  t^{\mathrm{h,h}}_{1,3} &=& -\frac{271}{3538944}\,h^{8}+\frac{3438923}{5308416000}\,h^{10}\nonumber\\
                       &&-\frac{3441481980911}{917294284800000}\,h^{12}\nonumber\\
  t^{\mathrm{h,v}}_{\frac{3}{2},\frac{5}{2}} &=& -\frac{271}{1179648}\,h^{8}+\frac{29122303}{21233664000}\,h^{10}\nonumber\\
                       &&-\frac{7447108569091}{1223059046400000}\,h^{12}\nonumber\\
  t^{\mathrm{h,h}}_{2,2} &=& -\frac{271}{1179648}\,h^{8}+\frac{32991041}{21233664000}\,h^{10}\nonumber\\
                       &&-\frac{12599471323}{1791590400000}\,h^{12}\nonumber\\
  t^{\mathrm{h,v}}_{\frac{5}{2},\frac{3}{2}} &=& -\frac{271}{1179648}\,h^{8}+\frac{29122303}{21233664000}\,h^{10}\nonumber\\
                       &&-\frac{7447108569091}{1223059046400000}\,h^{12}\nonumber\\
  t^{\mathrm{h,h}}_{3,1} &=& -\frac{271}{1179648}\,h^{8}+\frac{1683571}{1415577600}\,h^{10}\nonumber\\
                       &&-\frac{605405462489}{135895449600000}\,h^{12}\nonumber\\
  t^{\mathrm{h,v}}_{\frac{7}{2},\frac{1}{2}} &=& -\frac{271}{3538944}\,h^{8}+\frac{4822213}{10616832000}\,h^{10}\nonumber\\
                       &&-\frac{143095110961}{73383542784000}\,h^{12}\nonumber\\
  t^{\mathrm{h,h}}_{4,0} &=& -\frac{271}{3538944}\,h^{8}+\frac{138329}{530841600}\,h^{10}\nonumber\\
                       &&-\frac{2796720985871}{3669177139200000}\,h^{12}\nonumber\\
  t^{\mathrm{h,h}}_{0,5} &=& -\frac{168493133}{45864714240000}\,h^{12}\nonumber\\
  t^{\mathrm{h,v}}_{\frac{1}{2},\frac{9}{2}} &=& \frac{7579}{663552000}\,h^{10}-\frac{85039213657}{917294284800000}\,h^{12}\nonumber\\
  t^{\mathrm{h,h}}_{1,4} &=& \frac{7579}{663552000}\,h^{10}-\frac{12339018187}{91729428480000}\,h^{12}\nonumber\\
  t^{\mathrm{h,v}}_{\frac{3}{2},\frac{7}{2}} &=& \frac{7579}{165888000}\,h^{10}-\frac{1225804705393}{3210529996800000}\,h^{12}\nonumber\\
  t^{\mathrm{h,h}}_{2,3} &=& \frac{7579}{165888000}\,h^{10}-\frac{1476569703919}{3210529996800000}\,h^{12}\nonumber\\
  t^{\mathrm{h,v}}_{\frac{5}{2},\frac{5}{2}} &=& \frac{7579}{110592000}\,h^{10}-\frac{622709811499}{1070176665600000}\,h^{12}\nonumber\\
  t^{\mathrm{h,h}}_{3,2} &=& \frac{7579}{110592000}\,h^{10}-\frac{622709811499}{1070176665600000}\,h^{12}\nonumber\\
    t^{\mathrm{h,v}}_{\frac{7}{2},\frac{3}{2}} &=& \frac{7579}{165888000}\,h^{10}-\frac{1225804705393}{3210529996800000}\,h^{12}\nonumber\\
\end{eqnarray}
\begin{eqnarray}
  t^{\mathrm{h,h}}_{4,1} &=& \frac{7579}{165888000}\,h^{10}-\frac{975039706867}{3210529996800000}\,h^{12}\nonumber\\
  t^{\mathrm{h,v}}_{\frac{9}{2},\frac{1}{2}} &=& \frac{7579}{663552000}\,h^{10}-\frac{85039213657}{917294284800000}\,h^{12}\nonumber\\
  t^{\mathrm{h,h}}_{5,0} &=& \frac{7579}{663552000}\,h^{10}-\frac{11672061361}{229323571200000}\,h^{12}\nonumber\\
  t^{\mathrm{h,h}}_{0,6} &=& 0\nonumber\\
  t^{\mathrm{h,v}}_{\frac{1}{2},\frac{11}{2}} &=& -\frac{168493133}{91729428480000}\,h^{12}\nonumber\\
  t^{\mathrm{h,h}}_{1,5} &=& -\frac{168493133}{91729428480000}\,h^{12}\nonumber\\
  t^{\mathrm{h,v}}_{\frac{3}{2},\frac{9}{2}} &=& -\frac{168493133}{18345885696000}\,h^{12}\nonumber\\
  t^{\mathrm{h,h}}_{2,4} &=& -\frac{168493133}{18345885696000}\,h^{12}\nonumber\\
  t^{\mathrm{h,v}}_{\frac{5}{2},\frac{7}{2}} &=& -\frac{168493133}{9172942848000}\,h^{12}\nonumber\\
  t^{\mathrm{h,h}}_{3,3} &=& -\frac{168493133}{9172942848000}\,h^{12}\nonumber\\
  t^{\mathrm{h,v}}_{\frac{7}{2},\frac{5}{2}} &=& -\frac{168493133}{9172942848000}\,h^{12}\nonumber\\
  t^{\mathrm{h,h}}_{4,2} &=& -\frac{168493133}{9172942848000}\,h^{12}\nonumber\\
  t^{\mathrm{h,v}}_{\frac{9}{2},\frac{3}{2}} &=& -\frac{168493133}{18345885696000}\,h^{12}\nonumber\\
  t^{\mathrm{h,h}}_{5,1} &=& -\frac{168493133}{18345885696000}\,h^{12}\nonumber\\
  t^{\mathrm{h,v}}_{\frac{11}{2},\frac{1}{2}} &=& -\frac{168493133}{91729428480000}\,h^{12}\nonumber\\
  t^{\mathrm{h,h}}_{6,0} &=& -\frac{168493133}{91729428480000}\,h^{12}\nonumber
\end{eqnarray}

\section{Low-energy spectrum of the XXZ}
\label{app:low_energy_XXZ}

Notations are the same as in Appendix \ref{app:low_energy_TFIM}.

\subsection{One-magnon hopping amplitudes}
\label{app:sub:XXZ_hopping1}

\begin{eqnarray}
  t_{0,0} &=& 4-\frac{1}{3}\,\lambda^{2}+\frac{287}{864}\,\lambda^{4}-\frac{910529}{6220800}\,\lambda^{6}\nonumber\\
         &&+\frac{5792068288969}{57610828800000}\,\lambda^{8}\nonumber\\
  t_{1,1} &=& -\frac{1}{2}\,\lambda^{2}+\frac{11}{72}\,\lambda^{4}-\frac{2106349}{31104000}\,\lambda^{6}\nonumber\\
         &&+\frac{19716698831861}{307257753600000}\,\lambda^{8}\nonumber\\
  t_{2,0} &=& -\frac{1}{4}\,\lambda^{2}+\frac{1}{72}\,\lambda^{4}-\frac{752221}{20736000}\,\lambda^{6}\nonumber\\
         &&+\frac{19657769838433}{614515507200000}\,\lambda^{8}\nonumber\\
  t_{2,2} &=& -\frac{7}{192}\,\lambda^{4}-\frac{279853}{31104000}\,\lambda^{6}+\frac{1163045360221}{153628876800000}\,\lambda^{8}\nonumber
\end{eqnarray}
\begin{eqnarray}
  t_{4,0} &=& -\frac{7}{1152}\,\lambda^{4}-\frac{578213}{62208000}\,\lambda^{6}+\frac{14632795561}{5486745600000}\,\lambda^{8}\nonumber\\
  t_{3,3} &=& -\frac{251}{27648}\,\lambda^{6}+\frac{2563786289}{10241925120000}\,\lambda^{8}\nonumber\\
  t_{4,2} &=& -\frac{251}{36864}\,\lambda^{6}-\frac{17610038713}{20483850240000}\,\lambda^{8}\nonumber\\
  t_{5,1} &=& -\frac{251}{92160}\,\lambda^{6}-\frac{39527719681}{20483850240000}\,\lambda^{8}\nonumber\\
  t_{6,0} &=& -\frac{251}{552960}\,\lambda^{6}-\frac{3452274997}{2926264320000}\,\lambda^{8}\nonumber\\
  t_{4,4} &=& -\frac{327349}{159252480}\,\lambda^{8}\nonumber\\
  t_{5,3} &=& -\frac{327349}{199065600}\,\lambda^{8}\nonumber\\
  t_{6,2} &=& -\frac{327349}{398131200}\,\lambda^{8}\nonumber\\
  t_{7,1} &=& -\frac{327349}{1393459200}\,\lambda^{8}\nonumber\\
  t_{8,0} &=& -\frac{327349}{11147673600}\,\lambda^{8}\nonumber
\end{eqnarray}

\subsection{Two-magnon bound state hopping amplitudes}
\label{app:sub:XXZ_hopping2}

\begin{eqnarray}
  t^{\mathrm{h,h}}_{0,0} &=& 6-\frac{1}{2}\,\lambda^{2}-\frac{229}{2880}\,\lambda^{4}-\frac{10916263}{12441600}\,\lambda^{6}\nonumber\\
                       &&+\frac{77090072016313}{76814438400000}\,\lambda^{8}\nonumber\\
                         t^{\mathrm{h,h}}_{0,1} &=& -\frac{1}{3}\,\lambda^{2}+\frac{67}{216}\,\lambda^{4}-\frac{4985783}{24883200}\,\lambda^{6}\nonumber\\
                       &&+\frac{231067152851203}{263363788800000}\,\lambda^{8}\nonumber\\
  t^{\mathrm{h,v}}_{\frac{1}{2},\frac{1}{2}} &=& -\frac{1}{3}\,\lambda^{2}+\frac{89}{432}\,\lambda^{4}-\frac{943667}{2764800}\,\lambda^{6}\nonumber\\
                       &&+\frac{223581725823587}{263363788800000}\,\lambda^{8}\nonumber\\
  t^{\mathrm{h,h}}_{1,0} &=& \frac{1}{6}\,\lambda^{2}+\frac{11}{288}\,\lambda^{4}-\frac{584821}{6220800}\,\lambda^{6}\nonumber\\
                       &&+\frac{10646999914133}{26336378880000}\,\lambda^{8}\nonumber\\
  t^{\mathrm{h,h}}_{0,2} &=& -\frac{257}{1728}\,\lambda^{4}-\frac{123863}{921600}\,\lambda^{6}+\frac{14069731116209}{184354652160000}\,\lambda^{8}\nonumber\\
    t^{\mathrm{h,v}}_{\frac{1}{2},\frac{3}{2}} &=& -\frac{1}{12}\,\lambda^{2}-\frac{29}{864}\,\lambda^{4}-\frac{2229119}{49766400}\,\lambda^{6}\nonumber\\
                       &&+\frac{4078695604273}{15049359360000}\,\lambda^{8}\nonumber\\
  t^{\mathrm{h,h}}_{1,1} &=& -\frac{1}{12}\,\lambda^{2}-\frac{41}{576}\,\lambda^{4}-\frac{335161}{2488320}\,\lambda^{6}\nonumber\\
                       &&+\frac{3645776906291}{17557585920000}\,\lambda^{8}\nonumber\\
                         t^{\mathrm{h,v}}_{\frac{3}{2},\frac{1}{2}} &=& -\frac{1}{12}\,\lambda^{2}-\frac{29}{864}\,\lambda^{4}-\frac{2229119}{49766400}\,\lambda^{6}\nonumber\\
                       &&+\frac{4078695604273}{15049359360000}\,\lambda^{8}\nonumber
\end{eqnarray}
\begin{eqnarray}
 t^{\mathrm{h,h}}_{2,0} &=& -\frac{1}{12}\,\lambda^{2}-\frac{11}{432}\,\lambda^{4}-\frac{857713}{12441600}\,\lambda^{6}\nonumber\\
                       &&+\frac{688142204467}{9405849600000}\,\lambda^{8}\nonumber\\
  t^{\mathrm{h,h}}_{0,3} &=& -\frac{1}{1728}\,\lambda^{4}+\frac{970931}{62208000}\,\lambda^{6}+\frac{23063997226177}{921773260800000}\,\lambda^{8}\nonumber\\
  t^{\mathrm{h,v}}_{\frac{1}{2},\frac{5}{2}} &=& \frac{1}{432}\,\lambda^{4}-\frac{244283}{41472000}\,\lambda^{6}+\frac{30735370205093}{921773260800000}\,\lambda^{8}\nonumber\\
  t^{\mathrm{h,h}}_{1,2} &=& \frac{7}{1728}\,\lambda^{4}-\frac{12331}{15552000}\,\lambda^{6}+\frac{207639631734467}{1843546521600000}\,\lambda^{8}\nonumber\\
  t^{\mathrm{h,v}}_{\frac{3}{2},\frac{3}{2}} &=& \frac{1}{192}\,\lambda^{4}-\frac{270403}{4976640}\,\lambda^{6}+\frac{26215520575403}{307257753600000}\,\lambda^{8}\nonumber\\
  t^{\mathrm{h,h}}_{2,1} &=& \frac{1}{192}\,\lambda^{4}-\frac{428399}{24883200}\,\lambda^{6}+\frac{14013035309131}{184354652160000}\,\lambda^{8}\nonumber\\
  t^{\mathrm{h,v}}_{\frac{5}{2},\frac{1}{2}} &=& \frac{1}{432}\,\lambda^{4}-\frac{244283}{41472000}\,\lambda^{6}+\frac{30735370205093}{921773260800000}\,\lambda^{8}\nonumber\\
  t^{\mathrm{h,h}}_{3,0} &=& \frac{1}{1728}\,\lambda^{4}-\frac{7861}{6220800}\,\lambda^{6}+\frac{282505209937}{46088663040000}\,\lambda^{8}\nonumber\\
  t^{\mathrm{h,h}}_{0,4} &=& -\frac{7277}{12441600}\,\lambda^{6}+\frac{28585558937}{307257753600000}\,\lambda^{8}\nonumber\\
  t^{\mathrm{h,v}}_{\frac{1}{2},\frac{7}{2}} &=& -\frac{1}{3456}\,\lambda^{4}+\frac{298777}{248832000}\,\lambda^{6}+\frac{1744637644499}{245806202880000}\,\lambda^{8}\nonumber\\
  t^{\mathrm{h,h}}_{1,3} &=& -\frac{1}{3456}\,\lambda^{4}+\frac{77}{124416}\,\lambda^{6}-\frac{4258929380371}{921773260800000}\,\lambda^{8}\nonumber\\
  t^{\mathrm{h,v}}_{\frac{3}{2},\frac{5}{2}} &=& -\frac{1}{1152}\,\lambda^{4}+\frac{406687}{248832000}\,\lambda^{6}+\frac{1166929622801}{73741860864000}\,\lambda^{8}\nonumber\\
  t^{\mathrm{h,h}}_{2,2} &=& -\frac{1}{1152}\,\lambda^{4}+\frac{9367}{2488320}\,\lambda^{6}+\frac{31971087871}{7681443840000}\,\lambda^{8}\nonumber\\
  t^{\mathrm{h,v}}_{\frac{5}{2},\frac{3}{2}} &=& -\frac{1}{1152}\,\lambda^{4}+\frac{406687}{248832000}\,\lambda^{6}+\frac{1166929622801}{73741860864000}\,\lambda^{8}\nonumber\\
  t^{\mathrm{h,h}}_{3,1} &=& -\frac{1}{1152}\,\lambda^{4}+\frac{14579}{6912000}\,\lambda^{6}+\frac{18122969387783}{1843546521600000}\,\lambda^{8}\nonumber\\
  t^{\mathrm{h,v}}_{\frac{7}{2},\frac{1}{2}} &=& -\frac{1}{3456}\,\lambda^{4}+\frac{298777}{248832000}\,\lambda^{6}+\frac{1744637644499}{245806202880000}\,\lambda^{8}\nonumber\\
  t^{\mathrm{h,h}}_{4,0} &=& -\frac{1}{3456}\,\lambda^{4}-\frac{15667}{31104000}\,\lambda^{6}+\frac{1013784889699}{307257753600000}\,\lambda^{8}\nonumber\\
  t^{\mathrm{h,h}}_{0,5} &=& -\frac{217}{12441600}\,\lambda^{6}-\frac{12986330681}{92177326080000}\,\lambda^{8}\nonumber\\
  t^{\mathrm{h,v}}_{\frac{1}{2},\frac{9}{2}} &=& -\frac{991}{6220800}\,\lambda^{6}-\frac{1247791850591}{1843546521600000}\,\lambda^{8}\nonumber\\
  t^{\mathrm{h,h}}_{1,4} &=& -\frac{4181}{12441600}\,\lambda^{6}-\frac{2538082499891}{921773260800000}\,\lambda^{8}\nonumber\\
  t^{\mathrm{h,v}}_{\frac{3}{2},\frac{7}{2}} &=& -\frac{120691}{124416000}\,\lambda^{6}-\frac{99157135381}{3687093043200000}\,\lambda^{8}\nonumber\\
  t^{\mathrm{h,h}}_{2,3} &=& -\frac{20177}{15552000}\,\lambda^{6}-\frac{26346818453}{34139750400000}\,\lambda^{8}\nonumber\\
  t^{\mathrm{h,v}}_{\frac{5}{2},\frac{5}{2}} &=& -\frac{25489}{15552000}\,\lambda^{6}+\frac{1118638298843}{921773260800000}\,\lambda^{8}\nonumber\\
  t^{\mathrm{h,h}}_{3,2} &=& -\frac{25489}{15552000}\,\lambda^{6}+\frac{415281490573}{153628876800000}\,\lambda^{8}\nonumber\\
  t^{\mathrm{h,v}}_{\frac{7}{2},\frac{3}{2}} &=& -\frac{120691}{124416000}\,\lambda^{6}-\frac{99157135381}{3687093043200000}\,\lambda^{8}\nonumber\\
  t^{\mathrm{h,h}}_{4,1} &=& -\frac{39983}{62208000}\,\lambda^{6}+\frac{690474135503}{921773260800000}\,\lambda^{8}\nonumber\\
  t^{\mathrm{h,v}}_{\frac{9}{2},\frac{1}{2}} &=& -\frac{991}{6220800}\,\lambda^{6}-\frac{1247791850591}{1843546521600000}\,\lambda^{8}\nonumber
\end{eqnarray}
\begin{eqnarray}
  t^{\mathrm{h,h}}_{5,0} &=& \frac{217}{12441600}\,\lambda^{6}-\frac{17201780639}{184354652160000}\,\lambda^{8}\nonumber\\
  t^{\mathrm{h,h}}_{0,6} &=& \frac{11448487}{1755758592000}\,\lambda^{8}\nonumber\\
  t^{\mathrm{h,v}}_{\frac{1}{2},\frac{11}{2}} &=& -\frac{217}{24883200}\,\lambda^{6}-\frac{32024937673}{737418608640000}\,\lambda^{8}\nonumber\\
  t^{\mathrm{h,h}}_{1,5} &=& -\frac{217}{24883200}\,\lambda^{6}-\frac{12649730881}{368709304320000}\,\lambda^{8}\nonumber\\
  t^{\mathrm{h,v}}_{\frac{3}{2},\frac{9}{2}} &=& -\frac{217}{4976640}\,\lambda^{6}-\frac{158928910813}{737418608640000}\,\lambda^{8}\nonumber\\
  t^{\mathrm{h,h}}_{2,4} &=& -\frac{217}{4976640}\,\lambda^{6}-\frac{10506217709}{92177326080000}\,\lambda^{8}\nonumber\\
  t^{\mathrm{h,v}}_{\frac{5}{2},\frac{7}{2}} &=& -\frac{217}{2488320}\,\lambda^{6}-\frac{154367383127}{368709304320000}\,\lambda^{8}\nonumber\\
  t^{\mathrm{h,h}}_{3,3} &=& -\frac{217}{2488320}\,\lambda^{6}-\frac{7497652279}{20483850240000}\,\lambda^{8}\nonumber\\
  t^{\mathrm{h,v}}_{\frac{7}{2},\frac{5}{2}} &=& -\frac{217}{2488320}\,\lambda^{6}-\frac{154367383127}{368709304320000}\,\lambda^{8}\nonumber\\
  t^{\mathrm{h,h}}_{4,2} &=& -\frac{217}{2488320}\,\lambda^{6}-\frac{104001673121}{184354652160000}\,\lambda^{8}\nonumber\\
  t^{\mathrm{h,v}}_{\frac{9}{2},\frac{3}{2}} &=& -\frac{217}{4976640}\,\lambda^{6}-\frac{158928910813}{737418608640000}\,\lambda^{8}\nonumber\\
  t^{\mathrm{h,h}}_{5,1} &=& -\frac{217}{4976640}\,\lambda^{6}-\frac{101559204737}{368709304320000}\,\lambda^{8}\nonumber\\
  t^{\mathrm{h,v}}_{\frac{11}{2},\frac{1}{2}} &=& -\frac{217}{24883200}\,\lambda^{6}-\frac{32024937673}{737418608640000}\,\lambda^{8}\nonumber\\
  t^{\mathrm{h,h}}_{6,0} &=& -\frac{217}{24883200}\,\lambda^{6}-\frac{5743189}{1280240640000}\,\lambda^{8}\nonumber\\
  t^{\mathrm{h,h}}_{0,7} &=& -\frac{421}{5573836800}\,\lambda^{8}\nonumber\\
  t^{\mathrm{h,v}}_{\frac{1}{2},\frac{13}{2}} &=& \frac{2696353}{1755758592000}\,\lambda^{8}\nonumber\\
  t^{\mathrm{h,h}}_{1,6} &=& \frac{5260091}{1755758592000}\,\lambda^{8}\nonumber\\
  t^{\mathrm{h,v}}_{\frac{3}{2},\frac{11}{2}} &=& \frac{15978763}{548674560000}\,\lambda^{8}\nonumber\\
  t^{\mathrm{h,h}}_{2,5} &=& \frac{103416781}{2926264320000}\,\lambda^{8}\nonumber\\
  t^{\mathrm{h,v}}_{\frac{5}{2},\frac{9}{2}} &=& \frac{896000347}{8778792960000}\,\lambda^{8}\nonumber\\
  t^{\mathrm{h,h}}_{3,4} &=& \frac{965398397}{8778792960000}\,\lambda^{8}\nonumber\\
     t^{\mathrm{h,v}}_{\frac{7}{2},\frac{7}{2}} &=& \frac{1308306883}{8778792960000}\,\lambda^{8}\nonumber\\
  t^{\mathrm{h,h}}_{4,3} &=& \frac{1308306883}{8778792960000}\,\lambda^{8}\nonumber\\
  t^{\mathrm{h,v}}_{\frac{9}{2},\frac{5}{2}} &=& \frac{896000347}{8778792960000}\,\lambda^{8}\nonumber\\
  t^{\mathrm{h,h}}_{5,2} &=& \frac{275534099}{2926264320000}\,\lambda^{8}\nonumber\\
  t^{\mathrm{h,v}}_{\frac{11}{2},\frac{3}{2}} &=& \frac{15978763}{548674560000}\,\lambda^{8}\nonumber\\
  t^{\mathrm{h,h}}_{6,1} &=& \frac{201070073}{8778792960000}\,\lambda^{8}\nonumber\\
  t^{\mathrm{h,v}}_{\frac{13}{2},\frac{1}{2}} &=& \frac{2696353}{1755758592000}\,\lambda^{8}\nonumber
\end{eqnarray}
\begin{eqnarray}
  t^{\mathrm{h,h}}_{7,0} &=& \frac{421}{5573836800}\,\lambda^{8}\nonumber\\
  t^{\mathrm{h,h}}_{0,8} &=& 0\nonumber\\
  t^{\mathrm{h,v}}_{\frac{1}{2},\frac{15}{2}} &=& -\frac{421}{11147673600}\,\lambda^{8}\nonumber\\
  t^{\mathrm{h,h}}_{1,7} &=& -\frac{421}{11147673600}\,\lambda^{8}\nonumber\\
  t^{\mathrm{h,v}}_{\frac{3}{2},\frac{13}{2}} &=& -\frac{421}{1592524800}\,\lambda^{8}\nonumber\\
  t^{\mathrm{h,h}}_{2,6} &=& -\frac{421}{1592524800}\,\lambda^{8}\nonumber\\
  t^{\mathrm{h,v}}_{\frac{5}{2},\frac{11}{2}} &=& -\frac{421}{530841600}\,\lambda^{8}\nonumber\\
  t^{\mathrm{h,h}}_{3,5} &=& -\frac{421}{530841600}\,\lambda^{8}\nonumber\\
  t^{\mathrm{h,v}}_{\frac{7}{2},\frac{9}{2}} &=& -\frac{421}{318504960}\,\lambda^{8}\nonumber\\
  t^{\mathrm{h,h}}_{4,4} &=& -\frac{421}{318504960}\,\lambda^{8}\nonumber\\
  t^{\mathrm{h,v}}_{\frac{9}{2},\frac{7}{2}} &=& -\frac{421}{318504960}\,\lambda^{8}\nonumber\\
  t^{\mathrm{h,h}}_{5,3} &=& -\frac{421}{318504960}\,\lambda^{8}\nonumber\\
  t^{\mathrm{h,v}}_{\frac{11}{2},\frac{5}{2}} &=& -\frac{421}{530841600}\,\lambda^{8}\nonumber\\
  t^{\mathrm{h,h}}_{6,2} &=& -\frac{421}{530841600}\,\lambda^{8}\nonumber\\
  t^{\mathrm{h,v}}_{\frac{13}{2},\frac{3}{2}} &=& -\frac{421}{1592524800}\,\lambda^{8}\nonumber\\
  t^{\mathrm{h,h}}_{7,1} &=& -\frac{421}{1592524800}\,\lambda^{8}\nonumber\\
  t^{\mathrm{h,v}}_{\frac{15}{2},\frac{1}{2}} &=& -\frac{421}{11147673600}\,\lambda^{8}\nonumber\\
  t^{\mathrm{h,h}}_{8,0} &=& -\frac{421}{11147673600}\,\lambda^{8}\nonumber
\end{eqnarray}



\end{document}